\documentclass[preprint,12pt, 3p]{elsarticle}
 \usepackage{graphicx}
 \usepackage{amssymb}
 \usepackage{color}
 \usepackage{amsmath}
\usepackage{lineno}
\usepackage{appendix}

\newcommand{\denselist}{\setlength{\itemsep}{0cm} \setlength{\parskip}{0cm}}

 \journal{Astroparticle physics}

\begin{document}

 \begin{frontmatter}

 \title{Probing the evolution of the EAS muon content in the
        atmosphere with KASCADE-Grande}

\author[label2]{W.D. Apel}
\author[label3]{J.C.~Arteaga-Vel\'azquez\corref{cor1}}
\author[label2]{K.~Bekk}
\author[label1]{M. Bertaina}
\author[label2,label4]{J.~Bl\"umer\fnref{cor6}}
\author[label2]{H.~Bozdog}
\author[label5]{I.M.~Brancus}
\author[label1,label6]{E.~Cantoni\fnref{cor3}}
\author[label1]{A.~Chiavassa}
\author[label4]{F.~Cossavella\fnref{cor4}}
\author[label2]{K.~Daumiller}
\author[label7]{V.~de~Souza}
\author[label1]{F.~Di~Pierro}
\author[label2]{P.~Doll}
\author[label2]{R.~Engel}
\author[label8]{D.~Fuhrmann\fnref{cor5}}
\author[label5]{A.~Gherghel-Lascu}
\author[label2]{H.J.~Gils}
\author[label8]{R.~Glasstetter}
\author[label9]{C.~Grupen}
\author[label2]{A.~Haungs\corref{cor2}}
\author[label2]{D.~Heck}
\author[label10]{J.R.~H\"orandel}
\author[label2]{T.~Huege}
\author[label8]{K.-H.~Kampert}
\author[label2]{D.~Kang}
\author[label2]{H.O.~Klages}
\author[label2]{K.~Link}
\author[label11]{P.~{\L}uczak}
\author[label2]{H.J.~Mathes}
\author[label2]{H.J.~Mayer}
\author[label2]{J.~Milke}
\author[label5]{B.~Mitrica}
\author[label6]{C.~Morello}
\author[label2]{J.~Oehlschl\"ager}
\author[label12]{S.~Ostapchenko}
\author[label2]{T.~Pierog}
\author[label2]{H.~Rebel}
\author[label2]{M.~Roth}
\author[label2]{H.~Schieler}
\author[label2]{S.~Schoo}
\author[label2]{F.G.~Schr\"oder}
\author[label13]{O.~Sima}
\author[label5]{G.~Toma}
\author[label6]{G.C.~Trinchero}
\author[label2]{H.~Ulrich}
\author[label2]{A.~Weindl}
\author[label2]{J.~Wochele}
\author[label11]{J.~Zabierowski}

\address[label2]{Institut f\"ur Kernphysik, KIT - Karlsruher Institut f\"ur 
Technologie, Germany}
\address[label3]{Institute of Physics and Mathematics, Universidad Michoacana
 de San Nicol\'as de Hidalgo, Morelia, Mexico}
\address[label1]{Departimento di Fisica, Universit\`a degli Studi di Torino, Italy}
\address[label4]{Institut für Experimentelle Teilchenphysik, KIT - Karlsruher 
Institut für Technologie, Germany}
\address[label5]{Horia Hulubei National Institute of Physics and Nuclear 
Engineering, Bucharest, Romania}
\address[label6]{Osservatorio Astrofisico di Torino, INAF Torino, Italy}
\address[label7]{Universidade de S$\tilde{a}$o Paulo, Instituto de F\'{\i}sica de 
S\~ao Carlos, Brasil}
\address[label8]{Fachbereich Physik, Universit\"at Wuppertal, Germany}
\address[label9]{Department of Physics, Siegen University, Germany}
\address[label10]{Department of Astrophysics, Radboud University Nijmegen, The 
Netherlands}
\address[label11]{National Centre for Nuclear Research, Department of  
Astrophysics, Lodz, Poland}
\address[label12]{Frankfurt Institute for Advanced Studies (FIAS), Frankfurt am Main, Germany}
\address[label13]{Department of Physics, University of Bucharest, Bucharest, 
Romania}
\cortext[cor1]{Corresponding author: arteaga@ifm.umich.mx}
\cortext[cor2]{Spokesperson KASCADE-Grande:  andreas.haungs@kit.edu}
\fntext[cor6]{Now: Head of Division V at KIT -  Karlsruher Institut f\"ur 
Technologie, Germany}
\fntext[cor3]{Now at: Istituto Nazionale di Ricerca Metrologica, INRIM,
Torino, Italy}
\fntext[cor4]{Now at: DLR Oberpfaffenhofen, Germany}
\fntext[cor5]{Now at: University of Duisburg-Essen, Duisburg, Germany}

   \begin{abstract}
     
    \footnotesize         
    The evolution of the muon content of very high energy air showers (EAS) in the 
  atmosphere is investigated with data of the KASCADE-Grande observatory. For this 
  purpose, the muon attenuation length in the atmosphere is obtained to 
  $\Lambda_\mu = 1256 \, \pm 85 \, ^{+229}_{-232}(\mbox{syst})\, 
  \mbox{g/cm}^2$ from the experimental data for shower energies between $10^{16.3}$ 
  and  $10^{17.0} \, \mbox{eV}$. Comparison of this quantity with predictions of the 
  high-energy hadronic interaction models QGSJET-II-02, SIBYLL 2.1, QGSJET-II-04 and
  EPOS-LHC reveals that the attenuation of the muon content of measured EAS 
  in the atmosphere is lower than predicted. Deviations are, however, less 
  significant with the post-LHC models. The presence of such deviations seems to 
  be related to a difference between the simulated and the measured zenith angle 
  evolutions of the lateral muon density distributions of EAS, which also causes
  a discrepancy between the measured absorption lengths of the density of shower 
  muons and the predicted ones at large distances from the EAS core. The studied 
  deficiencies show that all four considered hadronic interaction models fail to 
  describe consistently the zenith angle evolution of the muon content of EAS in 
  the aforesaid energy regime.
  \end{abstract}


   \end{frontmatter}

  \section{Introduction}

   Extensive air showers (EAS) are cascades of secondary particles produced by multiple 
 particle reactions and decays in the atmosphere. These processes are triggered by  
 collisions of very high energy cosmic rays with the nuclei of the atmosphere. With 
 sophisticated air-shower detectors, the properties of the EAS  are measured  with the 
 aim of learning about the origin and physics of the parent cosmic rays, a task that it
 is often done by comparing the EAS data with Monte Carlo simulations. Critical elements 
 of these simulations are the hadronic interaction models, which rely on physical 
 extrapolations of accelerator measurements taken at lower energies \cite{MCOstap}. 
 The distinct phenomenological treatments employed in the models 
 and the uncertainties of the experimental input data lead to considerable differences 
 in the predictions of relevant EAS parameters at high energies \cite{MCOstap, MCPierog}, 
 which introduce significant model uncertainties when assigning the energy and 
 identifying the nature of the primary particles from the EAS measurements (see 
 for example \cite{KASCADE-model1}). Hence, it is imperative to check the validity 
 of the hadronic interaction models employed in the study of cosmic rays.

  At very high energies and in the kinematical regime relevant for cosmic ray physics, 
 the performance of hadronic interaction models can be checked by comparing their EAS 
 predictions with the data of air-shower observatories. Differences between model 
 expectations and experimental data found in this way can 
 not only serve to constrain the validity of the models but also to point out some of 
 their deficiencies as a basis for future model improvements. For testing the validity 
 of hadronic interaction models, muons play a particular role. Muons are created in 
 non-electromagnetic decays of shower hadrons, such as charged pions and  kaons. Once 
 produced, muons decouple immediately from the air shower and travel almost in straight 
 lines to the detector with smaller attenuation than that for electromagnetic and 
 hadronic particles \cite{Cazon}. Studying muons becomes therefore a sensitive and 
 direct way to probe the hadronic physics \cite{Meurer} and to identify 
 possible deficiencies of hadronic interaction models \cite{AugerXmaxMuon1, 
 AugerMuonExcess2}.
 
  In this regard, the present work aims to test the predictions of the high-energy
 hadronic interaction models QGSJET-II-02 \cite{qgs}, SIBYLL 2.1 \cite{sibyll}, 
 EPOS-LHC\cite{Pierog2013b} and QGSJET-II-04 \cite{qgs04} on the zenith-angle 
 dependence of the muon number in EAS.  The study is performed by measuring the 
 attenuation length of muons in air showers using the constant intensity cut (CIC) 
 method \cite{CIC} and by comparing the results with model predictions. The EAS 
 data were collected with the KASCADE-Grande observatory \cite{KG} during the 
 period from December 2003 to October 2011. 
 
 The paper is structured as follows: In section \ref{Setup} a brief description
 of the experimental KASCADE-Grande setup and the accuracy of the shower reconstruction
 at the observatory are presented as well as a short description of the selection cuts
 employed in the study. Then, in section \ref{MC}, the characteristics of the
 Monte Carlo data sets employed for the current investigation are described and the
 high-energy hadronic interaction models investigated in this study are briefly
 reviewed. The analyses  employed to test the hadronic interaction
 models are presented in detail in sections \ref{lambdamuon} and \ref{density}.
 The discussions of the results are reserved for section \ref{Discussion}.
 Section 7 \ref{Consequences} contains a brief account of the implications of the results for the
 features of the hadronic interaction models. In section \ref{Conclusions}, the
 conclusions of the present research are summarized. Finally, the statistical
 and systematic errors for our  results are listed and discussed in the
 appendices.

  \section{The KASCADE-Grande observatory}
  \label{Setup}

  \subsection{Experimental set-up} 
      
\begin{figure}[!t]
 \centering
 \includegraphics[width=3.2in]{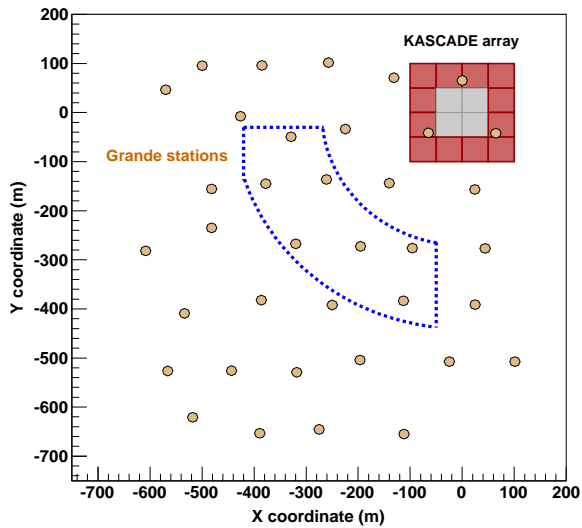}
  \caption{Layout of the KASCADE and the Grande arrays.  
	The circles mark the positions of the 37 Grande detector stations, while the 
  squares indicate the location of the 16 clusters in which were organized the 
  KASCADE detectors. The 12 outer clusters of the KASCADE array 
  housed 192 shielded plastic scintillator stations used for measurements
  of $N_\mu$. The dotted contour defines the area selected for the present analysis.}
 \label{KG_array}
\end{figure}

   The KASCADE-Grande experiment \cite{KG} was an air-shower array 
  devoted to study the energy spectrum and composition of cosmic rays with energies 
  between $E = 10^{16}$ and $10^{18} \, \mbox{eV}$, corresponding to center of 
  mass energies in the range of $\sqrt{s_{pp}} \approx 4.3 \,$ to  $43.3 \, \mbox{TeV}$. 
  The observatory was installed at the site of the KIT Campus North ($49.1^{\circ}$ N, 
  $8.4^{\circ}$ E, $110 \, \mbox{m}$ a.s.l.), Germany, and consisted of two 
  independent detector subsystems, the Grande and KASCADE arrays \cite{KG}. The 
  former was composed of a $700 \times 700 \, \mbox{m}^2$ grid of $37$ scintillator 
  stations regularly separated by an average distance of $137 \, \mbox{m}$ 
  (see fig.~\ref{KG_array}) and the latter, by a smaller and more compact 
  array of $252$ shielded and unshielded scintillation detectors spaced 
  every $13 \, \mbox{m}$ over a regular grid of $200 \times 200 \, \mbox{m}^2$ 
  overall surface. The Grande array provided ground measurements of the total 
  number of charged particles ($E > 3 \, \mbox{MeV}$), $N_{ch}$, at the EAS front, 
  while the KASCADE array was used to measure the corresponding total number of 
  muons ($E_\mu > 230 \, \mbox{MeV}$), $N_\mu$, among other observables. A more 
  detailed description of the KASCADE-Grande facility can be found in \cite{KG}. 
  
  \subsection{EAS reconstruction}
 
   Air shower reconstruction in KASCADE-Grande proceeds event-by-event by means of an 
  iterative algorithm and a careful modeling of the EAS front \cite{KG}. $N_{ch}$ is 
  estimated solely from the Grande data, while $N_{\mu}$ is derived from the 
  $\mu$-measurements of the KASCADE array. For the estimation  of $N_{ch}$  a 
  maximum-log-likelihood fit of a modified NKG lateral distribution function (LDF) 
  \cite{LDFNKG} is carried out using the densities of charged particles measured by 
  the Grande array for the event.

   For the estimation of $N_{\mu}$, in a first step, a calculation of the number of 
  muons detected in each KASCADE shielded station is performed. This is 
  accomplished by applying  a conversion function (LECF) to the energy deposit recorded in 
  each muon detector, whose main parameters have a negligible dependence
  on the shower size and the hadronic interaction model \cite{KG}. In the second and 
  last step, the total number of muons in the EAS is estimated with the maximum 
  likelihood technique by fitting a Lagutin-Raikin lateral distribution function 
  with a fixed shape \cite{LDFLR} to the data on the number of penetrating particles 
  registered by the KASCADE detectors:
  \begin{equation}
     \rho_\mu(r) = N_\mu \cdot \frac{0.28}{r_0^2} \left( \frac{r}{r_0}\right)^{p_1}
     \left( 1 + \frac{r}{r_0}\right)^{p_2}
     \left( 1 + \left( \frac{r}{10\cdot r_0}\right)^2 \right)^{p_3}
   \label{eq1}     
  \end{equation} 
  with $r$ the radial distance from the EAS core measured at the shower 
  plane. The parameters of the above equation are $p_1=-0.69$ , $p_2 = -2.39$, 
  $p_3= -1.0$ and  $r_0 = 320 \, \mbox{m}$ \cite{KG}. They were obtained calibrating 
  the function with the results of CORSIKA/QGSJET-01 simulations, in particular, by 
  averaging the fits to simulated protons and iron nuclei with energies of $10^{16}$ 
  and $10^{17} \, \mbox{eV}$. Fixing the shape of the muon lateral distribution 
  obeys to the limited statistics available from the muon detectors. If relaxing 
  this constraint on the LDF shape, the fit becomes unstable. 

     The resolution achieved by the whole fitting procedure is $\lesssim 15$\, \% 
  for $N_{ch}$ and  $\lesssim 25$\, \% for $N_{\mu}$. The first value was estimated
  in a model independent way \cite{KG} and the second one, from MC simulations 
  using the models under study (see \ref{appNmuCF}). For the upcoming 
  analysis, in order to improve the accuracy of the  muon number and consequently 
  on the determination of the muon attenuation length, $N_\mu$ was corrected for 
  experimental and reconstruction effects using a correction function (c.f. 
  \ref{appNmuCF}). The latter was built from MC simulations based on the 
  QGSJET-II-02 model. The choice of the MC model is not relevant for this task, 
  because the correction is almost independent of the high-energy hadronic 
  interaction model. After corrections, the mean $N_\mu$ systematic errors are 
  reduced to $\lesssim 10\, \%$ with a weak dependence on the core position, the 
  shower size, the muon size and the shower zenith angle in the full efficiency
  regime and, in particular, on the selected data sample.

 \subsection{Selection cuts and description of the data}
  \label{Dataset}

  Several selection cuts were developed in order to reduce the effect of 
  systematic uncertainties on the reconstructed shower observables, mainly on 
  $N_\mu$. These selection criteria were applied indistinctly to experimental 
  data and simulated events.
  
  First, selected events had to satisfy a $7/7$ Grande hardware trigger (six
  of a hexagonal shape and the central one) and activate more than 11
  Grande stations from a minimum number of 36 working Grande stations.
  Besides, all KASCADE detector clusters were required to be in operation during
  the data acquisition of the events. The quality of the reconstruction of the
  selected sample was assured by considering only events that passed successfully
  the standard reconstruction procedure of KASCADE-Grande. In addition, the
  selection for the present analysis included only events with their cores located
  at a distance between $270$ and $440 \, \mbox{m}$ from the KASCADE center and
  within a central area of $8 \times 10^{4} \, \mbox{m}^{2}$ inside the Grande 
  array (c.f. fig. \ref{KG_array}). With this cut not only  edge effects were 
  avoided but also a significant reduction of the $N_\mu$ systematic uncertainties 
  was achieved. In particular, events with a large contribution from the 
  electromagnetic punch-through effect were  eliminated. Showers with zenith angles
  greater than $40^{\circ}$ were also removed as they have a large pointing error.
  A further constraint on the data was set by introducing the limit $\log_{10} 
  N_\mu > 4.6$ on the reconstructed (not corrected yet) muon number for EAS. 
  This cut helped to discard a number of events below the efficiency threshold 
  irrelevant for the present analysis. 

   After these selection cuts, the full trigger and  reconstruction efficiency of 
  the KASCADE-Grande experiment is achieved at threshold energies around 
  $\log_{10} (E/\mbox{GeV}) = 7.00 \pm 0.20$ and corrected muon numbers 
  $\log_{10} N_\mu = 5.00 \pm 0.20$, according to MC simulations. The small 
  uncertainties originate from the unknown primary composition, the arrival direction 
  and the hadronic interaction model involved. For the selected events, the mean 
  core and pointing resolutions of KASCADE-Grande are better than $8 \, \mbox{m}$ 
  and $0.4^{\circ}$, respectively, and are almost independent of the radial distance 
  to the KASCADE array. The application of the selection criteria to the KASCADE-Grande 
  data yielded a data set with 2,744,950 shower events corresponding to a total 
  exposure of $2.6 \times 10^{12} \, \mbox{m}^2 \cdot \mbox{s} \cdot \mbox{sr}$.

  \section{Monte Carlo simulations}
  \label{MC}

   MC data were generated using simulations of the EAS development and of the response 
  of the detectors of the KASCADE-Grande array. In order to simulate the EAS 
  evolution in the atmosphere, the CORSIKA code \cite{CORSIKA} was used without employing 
  the thinning algorithm. The U.S. standard atmosphere model as parameterized by 
  J. Linsley (c.f. \cite{CORSIKA} and references therein) was employed, as  the mean of 
  the local RMS air pressure values at the site of the KASCADE-Grande observatory is 
  close to the magnitude predicted by the abovementioned model \cite{DanielThesis}. 

   The physics of the hadronic interactions was simulated using Fluka \cite{fluka} at 
  low energies ($E_h < 200 \, \mbox{GeV}$) combined with QGSJET-II-02,  SIBYLL 2.1,
  QGSJET-II-04 and EPOS-LHC as different alternatives to model the 
  high energy regime. MC showers were generated for the KASCADE-Grande location and 
  followed until they reach the detector level. The CORSIKA output was injected in a 
  GEANT 3.21 \cite{Geant} based code, where the response of the KASCADE-Grande 
  components were simulated in full detail and stored in data files, which have 
  the same format as the experimental data. The MC files were then processed with 
  the same KASCADE-Grande reconstruction program that was applied to the measured 
  data. This way, systematic uncertainties owing to the use of different reconstruction 
  algorithms were avoided. 

    The energy spectrum of the events in the MC data sets ranges from $10^{16}$  
  until $10^{18} \, \mbox{eV}$ and follows an $E^{-2}$ law. However, weights had 
  to be introduced and  applied to the MC data to simulate more realistic spectra 
  (see, for example, \cite{ KGPRD, KGNchNmu}) with $\gamma = -2.8, -3, -3.2$.
  Regarding the spatial distribution of the MC events, they are isotropically 
  distributed  and their cores at ground are homogeneously 
  scattered over the full array. Shower simulations are done up to zenith angles 
  of $42^{\circ}$ with no restriction for the azimuthal angle. Concerning composition, 
  MC data contain individual sets for different representative primaries: hydrogen 
  (H), helium (He), carbon (C), silicon (Si) and iron (Fe) nuclei, with roughly the 
  same statistics. An additional data set for each interaction model was also 
  included simulating a mixed composition scenario,  where the above elements are 
  present in equal abundances. The final QGSJET-II-02 data set with the five
  primaries contains $1.9$ million events, while the corresponding data files for
  the other models comprise roughly $1.2$ million events for SIBYLL 2.1,  $1.3$
  million events for QGSJET-II-04 and $2.2$ million events for EPOS-LHC.
  	
  Several differences are expected among the predictions of the various  
  hadronic interactions models for the KASCADE-Grande energy range at the altitude 
  of the observatory. Comparative studies performed for KASCADE-Grande showed that 
  QGSJET-II-02 produces a lower muon content in vertical EAS than the most recent 
  models QGSJET-II-04 and EPOS-LHC, but more muons than  SIBYLL 2.1 (e.g., at 
  $E \sim 10^{17} \, \mbox{eV}$, they amount to $\approx 13 \, \%$ and $21 \, \%$ 
  for the first two cases, respectively, and to $7\, \%$ for the last one). On the 
  other hand, it was found that QGSJET-II-02 predictions for the $N_\mu/N_{ch}$ ratio 
  in vertical showers are  smaller than the corresponding QGSJET-II-04 and EPOS-LHC 
  estimations ($18\, \%$ and $19\, \%$, respectively, at $E \sim 10^{17} \, \mbox{eV}$). 
  However, the QGSJET-II-02 ratios turned out to be almost equal to the SIBYLL 2.1 
  derived ones. The main reasons behind the muon enhancement in the current version of 
  QGSJET-II-04 are the larger $\pi^{\pm}$ production in pion-air interactions and the 
  harder pion spectra \cite{Ostap2013}. The latter is due to an increased forward 
  $\rho^{0}$ production in pion-nucleus collisions, compared to $\pi^0$ generation, 
  which enhances via the decay mode $\rho^{0} \rightarrow \pi^{+} \pi^{-}$ the 
  relative proportion of charged pions in EAS and leads to an increase of the 
  shower muon content \cite{Ostap2013}. In EPOS-LHC, an additional increase 
  of the muon production originates from an enhanced  production of baryon-antibaryon 
  pairs in pion-nucleus collisions, which  effectively increases the number of 
  hadron generations in the atmospheric nuclear cascades \cite{Pierog2013icrc}. 
  For more details concerning the models, predictions for other EAS observables, 
  and theoretical approaches see references \cite{ Ostap2013, Pierog2013icrc}.

  \section{The muon attenuation length}   
  \label{lambdamuon}

   We focus the present analysis to the calculation of the attenuation length of the number
  of shower muons in the atmosphere, $\Lambda_\mu$, as an appropriate physical quantity
  to study the evolution of the muon content of EAS in the atmosphere.
  This is an easy and direct way to compare the $N_\mu$ evolution
  observed in EAS with the predictions from MC simulations.
  In general, the EAS attenuation length is a quantity that measures the degree of 
  effective attenuation that a given air-shower component or observable undergoes in 
  the atmosphere. In particular, it is sensitive to the longitudinal development of 
  the EAS \cite{Grieder} and it is a\-ffected by the inelastic hadronic cross section 
  of the primary particle \cite{Hoerandel} and the underlying mechanisms of particle 
  production in the shower \cite{Pierog2013icrc}. The EAS attenuation length is, in 
  consequence, a physical quantity that encloses a large amount of information about the 
  generation and development of the air shower.   

   Alternative definitions exist for the EAS attenuation length depending of 
  the shower component and the applied experimental technique 
  (see for example \cite{Grieder,Hoerandel} and references therein). 
  Here, we will use the approach based on the Constant Intensity Cut (CIC) method 
  \cite{CIC}, as it is well-established and independent of the hadronic interaction 
  model. Pioneering work using the CIC method along with the $N_\mu$ data
  can be found  for exam\-ple in \cite{Akeno} and \cite{Sugar} (see also 
  \cite{Grieder} and references therein). The approach  has been exploited
  for a number of reasons at some EAS observatories, e.g. for the reconstruction 
  of the energy spectrum of cosmic rays \cite{Sugar}, the calculation of the
  $p-$Air cross section \cite{Akeno}, the test of hadronic interaction 
  models \cite{Sugar} and the extraction of $\Lambda_\mu$ \cite{Grieder, Akeno,
  Sugar}. However, in the latter case, the different experimental conditions,
  muon energy thresholds and EAS reconstruction methods of the observatories
  as well as the distinct column depths of the sites prevent us to compare those 
  early measurements of $\Lambda_\mu$ with that from the present paper. 
  
    The aim of the CIC method is to provide a way to relate data from 
  different zenith angles at roughly the same primary energies, without 
  any reference to MC simulations. This is achieved through the calculation of 
  attenuation curves at fixed shower rates. The CIC method is based on the  
  assumption that the arrival distribution of cosmic rays is isotropic
  so that the observed intensity of primary particles with the same energy 
  is independent of the zenith angle or the slant depth. 
  
\begin{figure}[!t]
 \centering
 \includegraphics[width=3.2in]{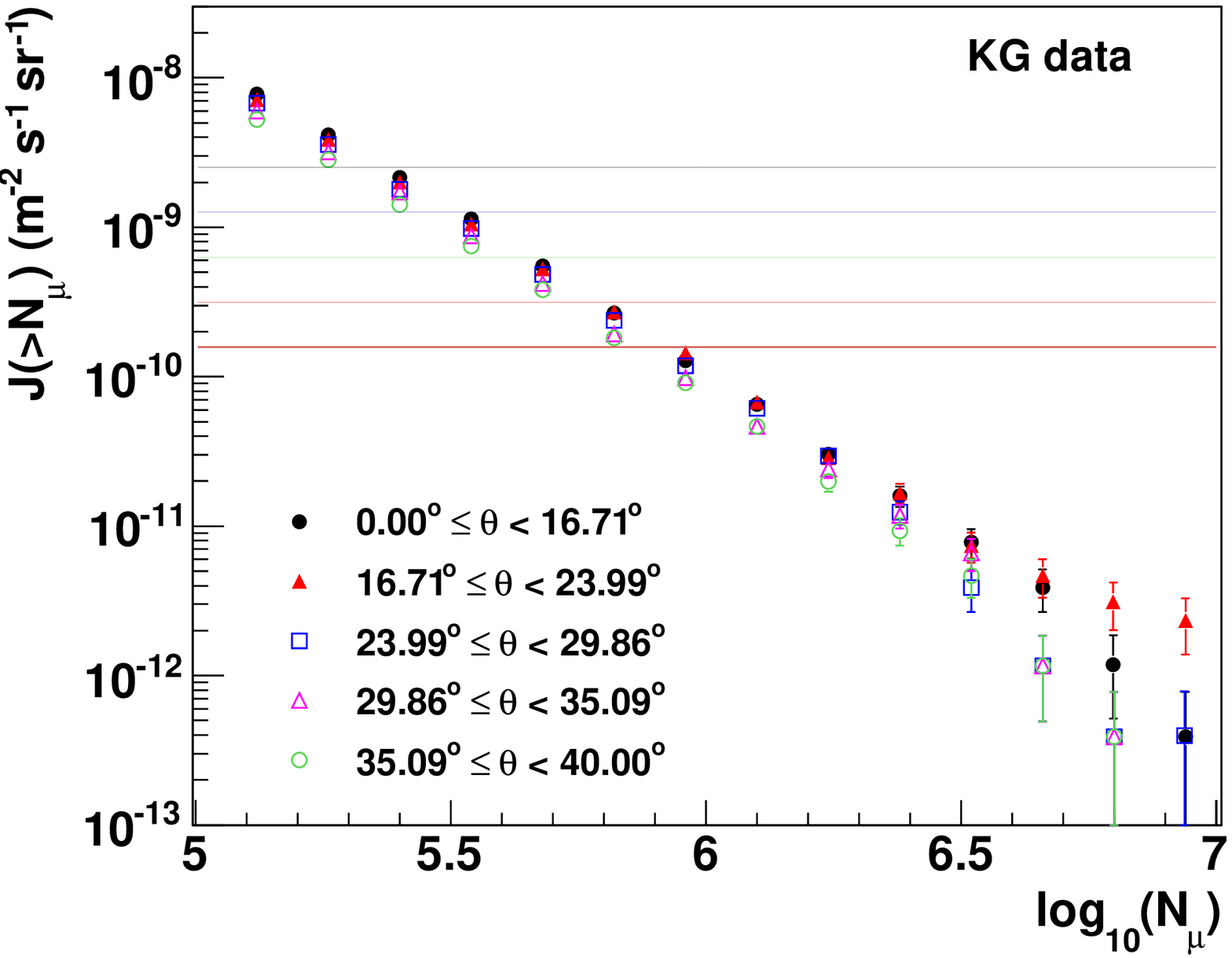}
 \includegraphics[width=3.2in]{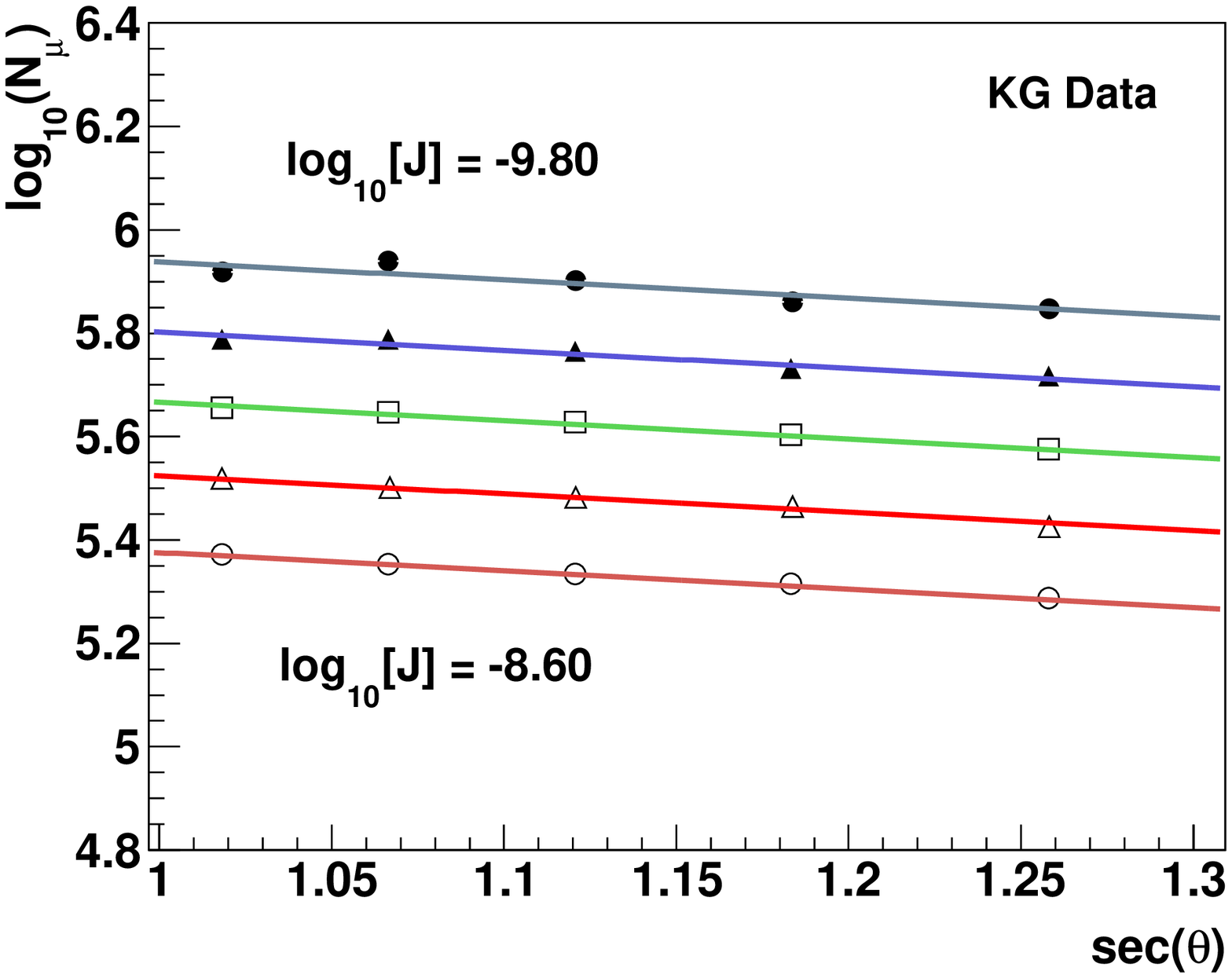}
  \caption{Left: Muon integral intensities for five zenith-angle intervals derived
  from the measurements with KASCADE-Grande, where the muon
  correction function is already applied. Error bars represent
  statistical uncertainties. The CIC employed are shown as 
  horizontal lines. Right: Muon attenuation curves obtained by applying the CIC 
	to the KASCADE-Grande integral spectra, $J_\mu$. The 
  cuts decrease from the bottom to the top in units of $\Delta \log_{10} 
  [J/(\mbox{m}^{-2} \cdot \mbox{s}^{-1} \cdot \mbox{sr}^{-1})] = -0.30$. Errors
  are smaller than the size of the symbols. They take into account 
  statistical uncertainties, errors from interpolation as well as  
  the correlation between adjacent points when interpolation was applied.
}
 \label{JmuandAttCurves}
\end{figure}

    In order to apply the CIC method, in the first instance, data were grouped 
  into five zenith-angle intervals with roughly the same aperture (see 
  fig.~\ref{JmuandAttCurves}, left). Then, for each angular bin the 
  corresponding integral muon intensity\footnote{Defined as the number of showers 
  detected above $N_\mu$ per unit solid angle, unit area and unit of time.} 
  $J(>N_\mu, \theta)$ is estimated  according to the following formula: 
  \begin{equation}
     J(>N_\mu, \theta) = \int_{N_\mu} \Phi(N_\mu, \theta) dN_\mu,  
  \label{eq2}
  \end{equation} 
  where $\Phi(N_\mu, \theta)$ represents the differential muon shower size
  spectrum.  

   Five cuts are applied on  $J(>N_\mu, \theta)$ at different 
  constant integral intensities in order to select showers with the
  same frequency rate at distinct zenith angles. This procedure is performed within the
  interval $\log_{10} N_\mu \approx [5.2, 6.0]$ of full efficiency and maximum 
  statistics as shown in fig.~\ref{JmuandAttCurves} (left).

   After that, the intersections of each cut with the integral spectra for the
  different angular bins are found\footnote{When necessary, interpolation
  between two adjacent points along the same intensity was applied by means of a 
  simple power-law expression.}. Then for each constant intensity cut, a
  muon attenuation curve $\log_{10} N_\mu(\theta)$ is built using the
  corresponding set of intersection points. The results are displayed on the 
  right plot of fig.~\ref{JmuandAttCurves} for all CIC cuts employed in the 
  study. These attenuation curves describe roughly the way in which the muon 
  content of an average EAS evolves in the atmosphere for different primary
  energies. Finally, in order to extract the value of the muon attenuation
  length ($\Lambda_\mu$) that best describes our data, a global fit via a
  $\chi^{2}$-minimization is applied to the set of attenuation curves  
  using 
  \begin{equation}
     N_\mu(\theta) = N_\mu^{\circ}\mbox{e}^{-X_0 \sec{\theta}/\Lambda_\mu},
  \label{eq3}
  \end{equation}
  with a common $\Lambda_\mu$, where $X_0 = 1022 \, \mbox{g}/\mbox{cm}^{2}$ 
  is the average atmospheric depth for vertical showers at the location of 
  the experiment and $N_\mu^{\circ}$ is a normalization parameter to be 
  determined for each attenuation curve. The analysis of both the MC and 
  measured data have shown that it is possible to use a single $\Lambda_\mu$ 
  for the entire $N_\mu$ range, as the standard deviation of the results  
  obtained when fitting individually the attenuation curves is smaller 
  than $\sim 3 \, \%$ in each case.
 \begin{figure}[!t]
 \centering
 \includegraphics[width=3.2in]{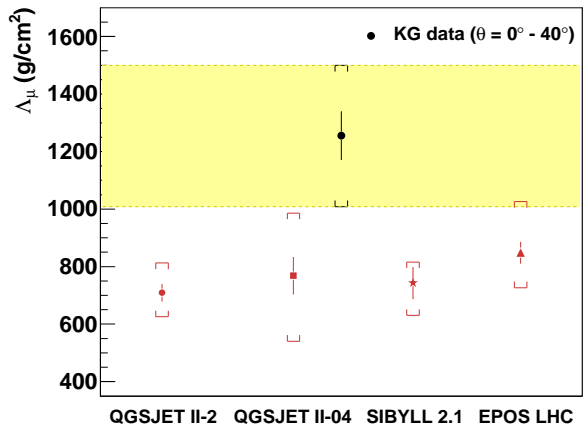}
  \caption{Muon attenuation lengths extracted from Monte Carlo 
  (points below shadowed area) and experimental data (upper black 
  circle). Error bars indicate statistical uncertainties, while
  the brackets represent the total errors (systematic plus statistical 
  errors added in quadrature). The shadowed band covers the total 
  uncertainty estimated for the experimental result.}
 \label{LambdaMuon}
\end{figure}

   The value of the muon attenuation length of EAS measured with the 
  KASCADE-Grande array is presented in table~\ref{tab1} and fig.~
  \ref{LambdaMuon} together with the values extracted from MC data by 
  applying the same analysis. The quoted values for $\Lambda_\mu$ 
  in case of MC data correspond to the predictions of different hadronic 
  interaction models tested under the assumption of a mixed composition 
  scenario with $\gamma = -3$. It must be mentioned that simulated data has 
  been normalized in such a way that MC muon size spectra for vertical showers are 
  equal to the measured one around $\log_{10}(N_\mu) = 5.5$. We should also 
  add that the mean primary energies of the shower events lying along the 
  attenuation curves shown in fig. \ref{JmuandAttCurves} (right) cover the energy 
  intervals 
  $\log_{10}(E/\mbox{GeV}) = [7.4, 8.0], [7.3, 7.9], [7.4, 8.0]$ and 
  $[7.3, 7.9]$ according to the QGSJET-II-02, QGSJET-II-04, SIBYLL 2.1 and
  EPOS-LHC models, respectively. These energy assignments were estimated 
  from $N_\mu$ using power-law formulas calibrated with MC data for each 
  zenith-angle interval. A primary cosmic ray spectrum characterized by a 
  mixed composition and a spectral index $\gamma = -3$ was used for the
  energy calibration. Returning to table~\ref{tab1}, results 
  are accompanied by their statistical and systematic uncertainties. 
  The experimental systematic error includes (a detailed discussion 
  can be found in \ref{appA}):
	\begin{itemize}
	\denselist
	\item uncertainty resulting from the CIC method;
	\item uncertainty owing to the size of the 
          zenith-angle intervals;
	\item uncertainty in the parameters of the muon correction function; 
	\item systematic bias of the corrected muon number and its model and
	      composition dependence;
    \item and uncertainties associated with the EAS core position relative to the 
          center of the KASCADE muon array.
	\end{itemize}
   In addition, the MC systematic error includes uncertainties associated with 
   the spectral index and primary composition.     
       
	 From fig.~\ref{LambdaMuon}, it is observed that the measured $\Lambda_\mu$  
   lies above the MC predictions. The deviations of the experimental value from 
   the MC expectations are shown in table~\ref{tab1} along with the confidence 
   levels (CL) for agreement with the model estimations. From both table~\ref{tab1} 
   and fig.~\ref{LambdaMuon}, it can be seen that the pre-LHC models QGSJET-II-02 
   and SIBYLL 2.1 show the largest discrepancies with deviations at the level
   of $+2.04 \, \sigma$ and $+1.99 \, \sigma$, respectively. The corresponding 
   CL's are $2.08 \, \%$ and $2.34 \, \%$ and indicate that the probability of 
   agreement between experiment and the expectations is low for these cases. On 
   the other hand, just slight discrepancies are found for the post-LHC models 
   QGSJET-II-04 and EPOS-LHC, with $+1.48 \, \sigma$ and $+1.34 \, \sigma$,
   respectively, which imply that both predictions are each in reasonably agreement 
   with the measured value with CL's of $7 \, \%$ and $9 \, \%$ , respectively. 
   In spite of this, however, the fact that the central 
   values of the QGSJET-II-04 and EPOS-LHC predictions are still below the 
   experimental data could mean that more work is still needed within these 
   post-LHC models to give also a precise description of the KASCADE-Grande air-shower 
   results (this seems
   to be the case as revealed by the complementary study performed in section 
   \ref{density}). 
   
   Potential sources of systematic errors which could explain the 
   observed deviation between the expectations and the measurement were studied and 
   are presented in \ref{appB}. Special attention was given to systematic shifts of 
   $\Lambda_\mu$ produced by 
   instrumental effects, reconstruction  procedures, EAS fluctuations and environmental 
   effects, e.g., the aging of the muon detectors, the core position and arrival 
   angle resolutions of the apparatus,  errors in the reconstructed  number of  
   muons from uncertainties in the deposited energy per KASCADE  shielded  detector, 
   the uncertainty in the $N_\mu$ correction function, fluctuations on the number of 
   registered particles per muon station, the evolution of the chemical composition 
   of cosmic rays in the energy range considered and the influence of local variations
   of the atmospheric temperature and pressure. However, the analyses have shown 
   that the disagreement on $\Lambda_\mu$ between MC predictions and the experimental 
   measurement can not be ascribed to any of the above sources. We also investigated 
   the uncertainty in the shape of the muon LDF employed with the EAS data. Here we 
   show that it contributes to the discrepancy, but it is not the leading effect. 
   Therefore, the observed difference very likely originates from deficiencies of 
   the muon predictions of the high-energy hadronic interaction models underlying 
   our studies.

   \begin{table}[t]
    \begin{center}
    \caption{Muon attenuation lengths extracted from Monte Carlo and experimental
    data. $\Lambda_\mu$ is presented along with their statistical and  systematic
    errors (in order of appearance). Also given are deviations (in units of $\sigma$) 
    of the measured $\Lambda_\mu$ from the predictions of different hadronic 
    interaction models. The one-tailed confidence levels (CL) that the measured 
    value is in agreement with the MC predictions are also presented.}
    \small
   \begin{tabular}{l|c|c|c|c|c}
   \hline 
    &  QGSJET-II-02 & QGSJET-II-04 & SIBYLL 2.1  & EPOS-LHC & KG data \\ 
   \hline
   &&&&\\
   {\bf $\Lambda_\mu$ $(\mbox{g/cm}^2)$} 
   &$709 \pm 30^{+99}_{-78}$ 
   &$768 \pm 65^{+208}_{-219}$ 
   &$743 \pm 56^{+47}_{-98}$  
   &$848 \pm 38^{+174}_{-115}$
   &$1256 \pm 85^{+229}_{-232}$\\ 
   &&&&\\
   \hline
	 Deviation ($\sigma$) & $+2.04$ & $+1.48$ & $+1.99$ & $+1.34$ & \\ 
   CL (\%)                & $2.08$  & $6.96$  & $2.34$  & $9.07$  & \\
   \hline
   \end{tabular}
   \label{tab1}
   \end{center}
  \end{table}

 	\begin{figure}[!t]
    \centering
    \includegraphics[width=3.2in]{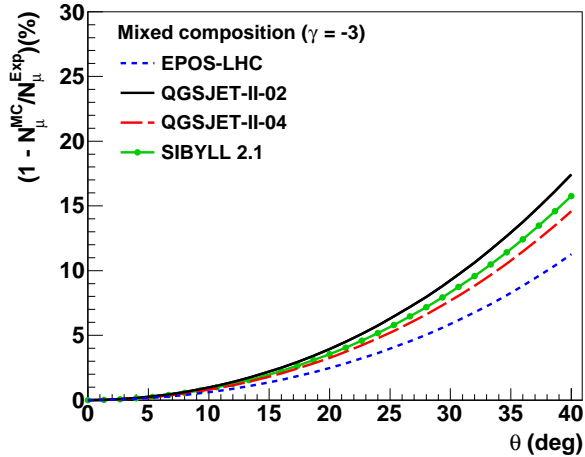}
    \caption{Relative differences in percentage between the measured muon content
    of EAS and the corresponding predictions of MC models under
    the assumption that the actual $N_\mu$ in vertical showers is well described
    by the models. Differences are presented as a function of the
    zenith angle according to equation (\ref{eq4}).}
    \label{NmuCFDeviationKG-MC}
   \end{figure} 
 	
      The fact that the experimental value of $\Lambda_\mu$ is greater than 
   the expected values from MC simulations implies that the observed air 
   showers attenuate more slowly in the atmosphere than the simulated 
   ones. It is difficult at this point to quantify the influence of such an 
   effect on the differences between the predicted and measured muon content 
   of air showers at large zenith angles. However, a naive estimation can be 
   done by assuming that for vertical showers the  MC predictions for the muon 
   number agree with the measured values at the same energy. Using equation 
   (\ref{eq3}), it is found that the $N_\mu$ differences, $\Delta_\mu$, expected 
   between measurements and MC predictions at different zenith angles, are given 
   by
   \begin{equation}
    \Delta_\mu = 1 - N_\mu^{MC}(\theta)/N_\mu^{Exp}(\theta) = 1 -  
     \mbox{e}^{-X_0 \cdot (\sec{\theta} - 1) \cdot 
     (1/\Lambda^{MC}_\mu  - 1/\Lambda^{Exp}_\mu)},
    \label{eq4}
   \end{equation}
   where the simulated attenuation curves have been normalized at $\theta = 0^{ \circ}$ 
   in such a way that $N_\mu^{MC}(0^{ \circ}) = N_\mu^{Exp}(0^{ \circ})$. Predictions 
   do not take into account systematic uncertainties from the reconstruction method, 
   experimental errors or the analysis technique. From fig.~\ref{NmuCFDeviationKG-MC} 
   it is observed that the $\Delta_\mu$ differences increase exponentially for 
   inclined showers becoming as high as $18\, \%$ at $\theta = 40^{ \circ}$. Note that 
   QGSJET-II-02 gives the highest differences due to its lower muon attenuation
   length (c.f. table~\ref{tab1}). On the contrary, the smallest differences are found 
   in case of EPOS-LHC. In general, the results shown in fig.~\ref{NmuCFDeviationKG-MC} 
   imply that a higher $N_\mu$ should be expected in data than in MC events for air 
   showers arriving at high zenith angles. Of course, it could also happen that both 
   measurements and predictions are in better agreement at high zenith angles, which 
   would suggest a smaller muon content for the actual vertical EAS in comparison 
   with simulations. To settle down the question a direct measurement of the shower 
   energy, independent of MC calibrations as much as possible, would be necessary.
   Unfortunately this is not the case for KASCADE-Grande, where the energy
   is estimated in a model dependent way from the measured EAS observables and has
   an uncertainty associated with the primary composition \cite{KGNchNmu}.

 \section{The muon absorption length}
  \label{density}

     To have a better understanding of the observed deviations and to
   avoid some of the sources of systematic uncertainties discussed above,
   we study now the zenith-angle evolution of the muon component of EAS 
   using the mean local density  of muons instead of the $N_\mu$ 
   observable for showers with about the same primary energy. 
   The quantity reflecting this evolution is the muon absorption length, 
   $\alpha_\mu$, also called the attenuation length of $\rho_\mu(r)$ 
   \cite{Ave01}. To proceed in  a model 
   independent way, the CIC method is applied again, however, on $N_{ch}$  
   in place of the muon number, since the former has a lower systematic 
   uncertainty and its observed zenith-angle evolution  is in better 
   agreement with the MC calculations. Besides, because using $N_{ch}$ as an 
   energy estimator provides a way to avoid possible systematic errors 
   associated with $N_\mu$ that might contribute to the discrepancy observed 
   on the muon content of EAS. The only drawback is that $N_{ch}$ is subject 
   to bigger shower fluctuations than $N_\mu$ at the same energy, which
   causes a reduction of the measured $\alpha_\mu$ for decreasing values 
   of the shower size. This effect is the result of a bias driven mainly 
   by the influence of shower-to-shower fluctuations of  $N_{ch}$  on the 
   EAS selection. In order to reduce it, only data with large $N_{ch}$ 
   were selected for the present study, in particular, with 
   $E \approx 10^{17} \, \mbox{eV}$.       
       
\begin{figure}[!t]
 \centering
 \includegraphics[width=3.2in]{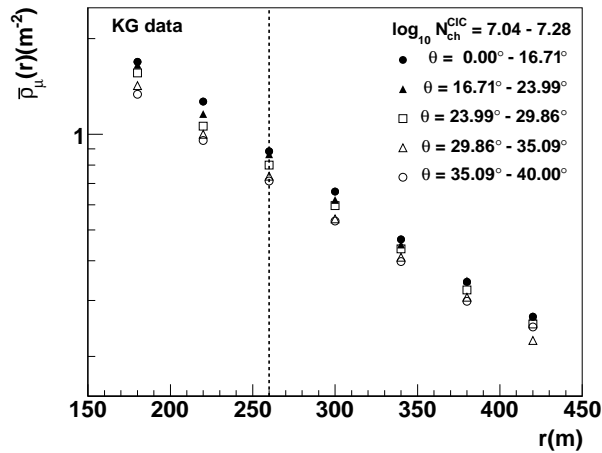}
 \caption{Mean lateral distributions of measured local muon densities 
 for several zenith-angle intervals and the shower size
 range $\log_{10} N_{ch}^{CIC}  = [7.04, 7.28]$. The vertical 
 line is an example of the cuts applied at a fixed 
 radius to extract the corresponding muon absorption 
 lengths. Statistical error bars of the data points are 
 smaller than the size of the markers.}
 \label{MuonDensitiesKG2}
\end{figure}

     Using the CIC method, $\Lambda_{ch}$ was estimated (see \ref{Lambdach}) 
   and afterwards employed to calculate the equivalent charged number of particles, 
   $N_{ch}^{CIC}$, at a zenith angle of reference, $\theta_{ref} = 22^{\circ}$
   (the mean of the zenith-angle distribution of experimental data). 
   This shower size observable was then used to select events in the interval 
   $\log_{10} N_{ch}^{CIC}  = [7.04, 7.28]$, roughly corresponding to the energy 
   region\footnote{In particular, for a mixed  composition assumption and  a 
   power-law  energy spectrum  $\propto E^{-3}$, the $N_{ch}^{CIC}$ intervals include 
   data with mean energy in the ranges of $\log_{10} (E/\mbox{GeV}) = [7.91, 8.14]$, 
   $[7.97, 8.20]$, $[7.95, 8.16]$ and $[7.89, 8.10]$ for QGSJET-II-02, QGSJET-II-04, 
   SIBYLL 2.1 and EPOS-LHC, respectively. Energy estimations were based in MC
   calibrated relations between the primary energy and the shower size for 
   $\theta = [21^{\circ}, 23^{\circ}]$.} from $\approx 10^{16.9}$ to $\approx 
   10^{17.2} \, \mbox{eV}$.  Events were further classified into five zenith
   angle intervals (with the same ranges used in the analysis of  $\Lambda_\mu$) 
   and within each of these bins, the mean muon densities at the shower plane, 
   $\bar{\rho}_\mu(r)$, were obtained. The procedure consists of dividing the shower 
   plane in concentric rings ($40 \, \mbox{m}$ width each) and then, for each $\theta$ 
   interval and radial bin, in dividing the total number of detected muons by the 
   corresponding sum of projected effective areas of the muon detectors registered 
   as active during the data taking of each selected event. 
   No corrections for atmospheric attenuation effects were included 
   when passing the muon data from the coordinate system of the detector to 
   that of the shower plane. The experimental results for the mean LDF of muons      
   within the above ranges are presented in fig. \ref{MuonDensitiesKG2}.

    To quantify $\alpha_\mu(r)$, absorption curves $\log_{10} \bar{\rho}_\mu 
   (r)$ vs $\sec(\theta)$ were further calculated. The curves were
   obtained from the $\bar{\rho}_\mu(r)$ distributions by applying several
   cuts at fixed distances $r$ from the EAS core at the shower plane (see fig.
   \ref{MuonDensitiesKG2}, for example). Cuts were applied in the interval 
   $r = [180 \, \mbox{m}, 380 \, \mbox{m}]$, where statistical fluctuations
   are low. For each  absorption curve, the muon absorption length, 
   $\alpha_\mu(r)$, was then estimated by fitting the data with the 
   following relation:
   \begin{equation}
     \bar{\rho}_\mu (r, \theta) = \bar{\rho}_\mu^{\circ}(r) 
     \mbox{e}^{-X_0 \sec{\theta}/\alpha_\mu(r)}, 
    \label{eq5}
   \end{equation}
   where $\bar{\rho}_\mu^{\circ}(r)$ is a normalization parameter.   

 \begin{figure}[!t]
 \centering
 \includegraphics[width=5.5in]{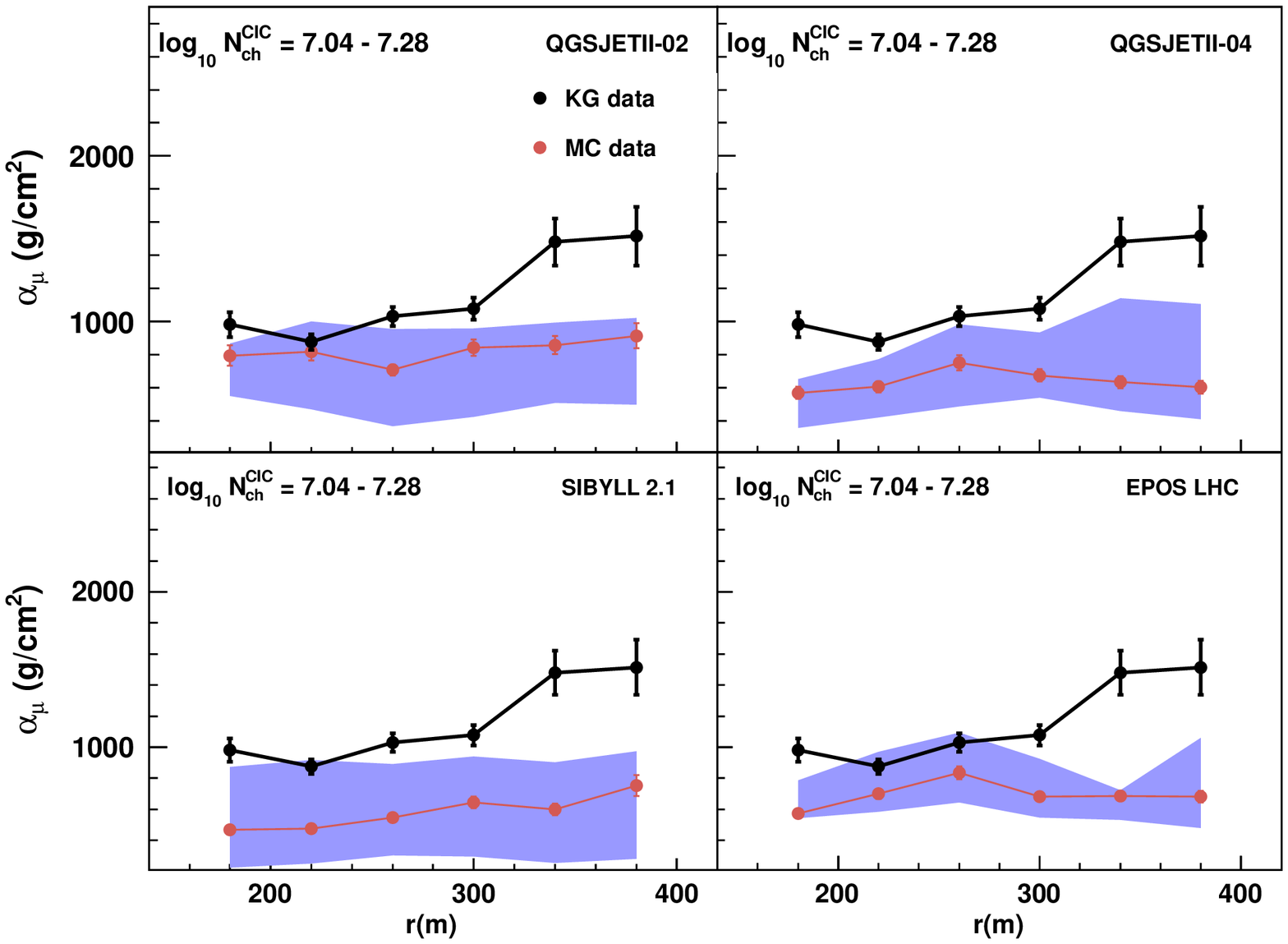}
 \caption{Muon absorption lengths for measured (KG label) and 
 simulated EAS data (MC label)  plotted against 
 the radial distance at the
 shower disk plane for the given $\log_{10} N_{ch}^{CIC}$
 interval. The error bands represent the
 systematic uncertainties due to composition and the spectral 
 index of the primary cosmic ray spectrum (see text). The
 error bars represent the uncertainties from the fits. 
 Note that, in some cases, for  MC data points the error bars 
 are smaller than the size of the markers.}
 \label{Results_absorption_New}
\end{figure}

    Fig. \ref{Results_absorption_New} shows the values of $\alpha_\mu$ 
   extracted from the KASCADE-Grande data for the chosen $N_{ch}^{CIC}$ interval 
   together with the predictions of MC simulations for different 
   hadronic interaction models. The MC values were calculated for a 
   mixed composition assumption and a primary spectral index of $\gamma=-3$. 
   The predicted $\alpha_\mu$ curves are accompanied by shadowed error bands
   that take into account the systematic errors due to both, composition and 
   spectral index uncertainties in the primary spectrum.
   The errors associated with the spectral index were obtained by repeating 
   the calculations with $\gamma=-2.8$ and $-3.2$, while the errors 
   due to composition were estimated by considering the distinct primary
   nuclei simulated in our MC data samples.  
   	  
 \begin{figure}[!t]
 \centering
 \includegraphics[width=6.8in]{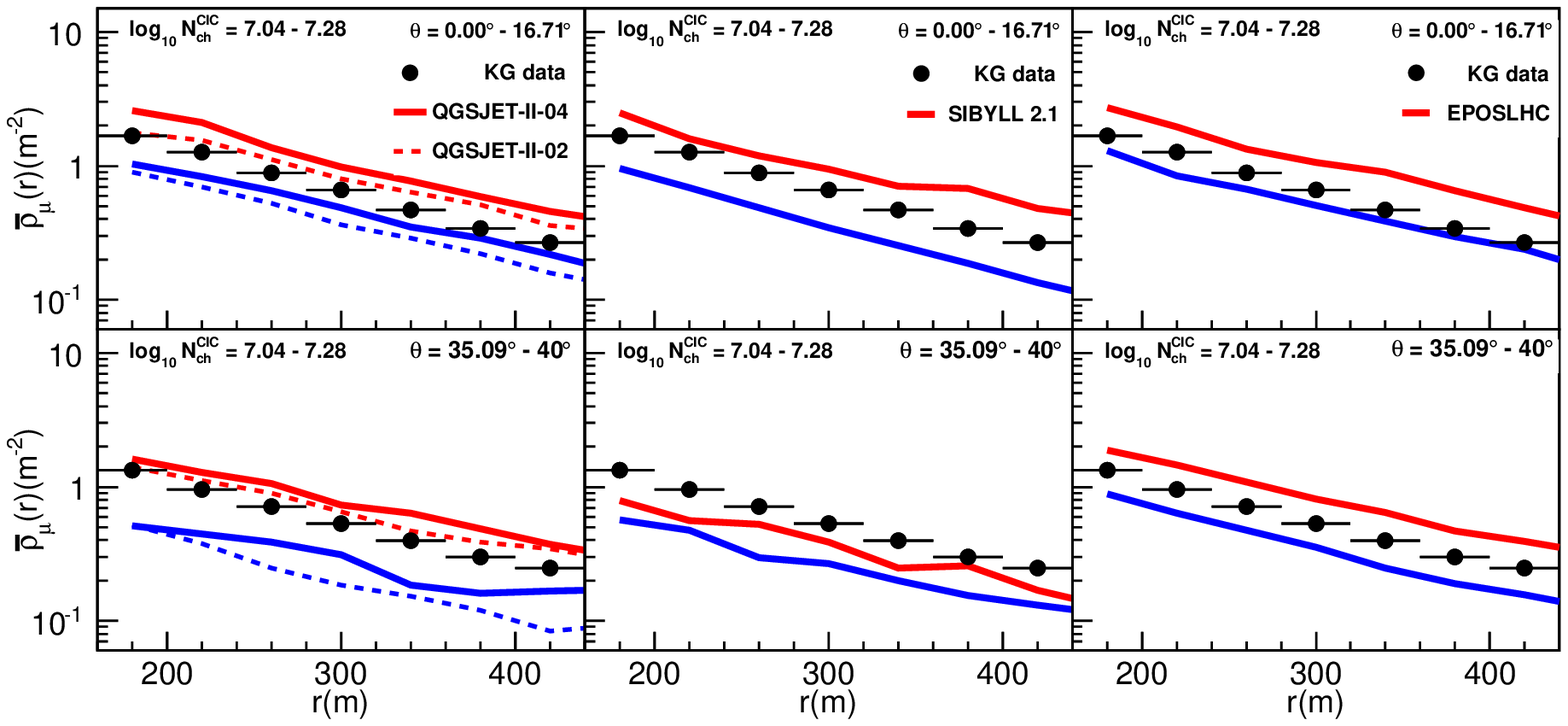}
\caption{Mean muon lateral distributions of EAS for the bin 
 $\log_{10} N_{ch}^{CIC}  = [7.04, 7.28]$ and two different 
 zenith-angle intervals. The solid points represent the
 experimental data and the lines the predictions from the models.
 Left column: QGSJET-II models; middle: SIBYLL 2.1; right: EPOS-LHC.
 For each model, results are shown for protons and iron nuclei 
 (lower and upper lines, respectively). Statistical error bars of the data points are 
 smaller than the size of the markers.}
 \label{DensitiesKG_vs_Models_zenith}
\end{figure}
	
	It is evident from fig.~\ref{Results_absorption_New} that the evolution 
   of the measured $\bar{\rho}_\mu(r)$ distributions in the atmosphere is not in 
   agreement with the expectations of the hadronic interaction models studied 
   in this work. We found that the measured $\alpha_\mu$ tends to stay above 
   the MC predictions and that there is only a marginal agreement between the 
   models and the experimental data  for radial distances closer to the shower core. 
   Fig. \ref{Results_absorption_New} shows that the differences between the 
   measurements and the model calculations rise with the lateral distance to the 
   core of the EAS. Strikingly, the $\Lambda_\mu$ parameter exhibits a similar 
   radial behavior as it was verified during the study of systematic errors 
   (see \ref{appA}) and in further analyses based on muon data around the EAS 
   core\footnote{We selected events with EAS cores within $58-250 \, \mbox{m}$ 
   from the center of the KASCADE array and applied the whole analysis described 
   in this paper to extract $\Lambda_\mu$ from the MC and the experimental data. 
   For QGSJET-II-02, we found a negligible variation of $\Lambda_\mu$ 
   with respect to the corresponding value of table \ref{tab1}, but 
   for the measured data  a reduction of almost $\approx 400 \, 
   \mbox{g}/\mbox{cm}^2$ was obtained, increasing the agreement 
   with model predictions.}. In consequence, we can conclude that the
   inconsistencies observed in the study of $\Lambda_\mu$ are still 
   present in the data for the local muon densities. Therefore, the 
   referred disagreements are not an artefact of the treatment of the 
   $N_\mu$ data or the way in which this parameter is estimated from the 
   particle densities at the muon detectors.
   
   Thus, in view of the above results, it seems entirely justifiable to say 
   that the discrepancies observed in the analysis of the local $\bar{\rho}_\mu(r)$ 
   distributions are the main responsible for the disagreement discovered 
   in the analysis of $\Lambda_\mu$. This asseveration was further supported 
   by additional tests carried out with Monte Carlo data (c.f.~\ref{appB}), 
   in which we observed that after increasing $\alpha_\mu$ in MC simulations 
   to reproduce the measured value, the experimental result of $\Lambda_\mu$ 
   can be recovered from the MC events.
  
    Here, it is important to add that despite the above deviations, the measured 
   muon densities for $\theta < 40^\circ$ along the corresponding CIC curve are 
   still bracketed by the estimations from the QGSJET-II-02, QGSJET-II-04 and 
   EPOS-LHC models for proton and iron nuclei, at least for the interval 
   $r = [180 \, \mbox{m}, 440 \, \mbox{m}]$. This is demonstrated in 
   fig.~\ref{DensitiesKG_vs_Models_zenith}. In contrast, for SIBYLL 2.1, 
   the situation is different, model predictions for proton and iron primaries do 
   not contain the measured data for inclined showers ($35.1^\circ \leq \theta 
   \leq 40^\circ$) within the shower size range $\log_{10} N_{ch}^{CIC} = 7.04-7.28$. 
   This result reveals an additional deficiency of the SIBYLL 2.1 model. However,
   it does not allow us to determine  whether the model  underestimates or 
   overestimates the muon content of EAS. The reasons are that, first, the 
   result depends on the reference angle that is used to find $N_{ch}^{CIC}$ 
   and, second, the energy calibration in KASCADE-Grande is model and composition 
   dependent.

  \section{Discussion of the measurements}
  \label{Discussion}
 
  The attenuation length of $N_\mu$ was measured at KASCADE-Grande for energies 
  between $10^{16.3}$ and $10^{17.0} \,\mbox{eV}$. The measured value is higher 
  than the predictions of QGSJET-II-02 and SIBYLL 2.1 but just exceeds slightly 
  the model calculations for EPOS-LHC and QGSJET-II-04 (see table \ref{tab1}). 
  The presence of such deviations was confirmed by the study of the 
  $\alpha_\mu(r)$ coefficients of the $\bar{\rho}_\mu$ distributions measured 
  locally at KASCADE-Grande around $E = 10^{17} \,\mbox{eV}$. This analysis showed 
  that the actual $\alpha_\mu(r)$ parameters become increasingly bigger 
  than the predicted MC values at large distances from the EAS core  
  (c.f. fig.~\ref{Results_absorption_New}). 
  The anomaly seems to be mainly associated to a bad description of the $\theta$ 
  dependence of the muon LDF's by the MC simulations (see \ref{appC5} and \ref{appC6}). 
  On the grounds of the above results, a general conclusion is derived that the 
  high-energy hadronic interaction models here analyzed can not describe consistently 
  all the muon data of EAS measured with the KASCADE-Grande array at different zenith 
  angles\footnote{Recently the post-LHC version of the SIBYLL model was released
  \cite{sibyll2.3}. The performance of this model at KASCADE-Grande is still
  under investigation. Results will be presented elsewhere.}.
      
    When extracting $\Lambda_\mu$ from the experimental data some input from the MC 
  models was unavoidable. First, through the lateral energy conversion function (LECF) 
  to  estimate the number of muons detected per muon station, then through the muon 
  LDF  employed to estimate $N_\mu$ and finally, through the muon correction function 
  introduced to correct $N_\mu$ for systematic biases. One may suppose that the 
  inclusion of such functions could invalidate the comparison between data and MC 
  predictions. Nevertheless, the model-experiment comparison of the EAS data is 
  completely justified, as we have processed and analysed both the experimental and 
  simulated events in identical ways using the same MC functions. Under the foregoing 
  procedure, however, it may become unclear whether the observed discrepancies are 
  due to the studied phenomenon or to a misleading description of the aforesaid 
  functions by the hadronic interaction models.

   The possibility that the MC based functions introduce the observed deviations
  in the  $\Lambda_\mu$ results seems to be weakened in view of the small model 
  dependence that these functions show (c.f.~\cite{KG} and \ref{appNmuCF}) and due 
  to the small variations that the relative systematic errors of $N_\mu$ exhibit 
  with the model (see fig.~\ref{MuonCF1}, left). These 
  kind of arguments are often invoked to validate some present studies (see, e.g. 
  \cite{AugerMuonExcess2}). However, one can argue that they do not constitute 
  a solid proof against the possibility being discussed. In this regard, it is 
  desirable to rely on additional analyses. For this reason, we have run 
  the complementary tests performed in section \ref{density} and \ref{appB}.
  As we have seen before, the former shows that anomalies are still present when 
  performing the analysis directly on the $\bar{\rho}_\mu$ data without any reference 
  to the muon LDF or the corresponding $N_\mu$ correction function (see fig.~ 
  \ref{Results_absorption_New}). While studies on \ref{appC5} have pointed out 
  that the experimental uncertainties on the shape of the muon LDF have not a 
  leading effect on the observed $\Lambda_\mu$ deviations. The tests however 
  did not deal with the muon LECF.
     
   The muon LECF correlates the energy losses by all particles in the KASCADE 
  shielded stations with the number of crossing muons. Therefore, if the
  real contribution from electrons, photons and hadrons is not well described
  by the models an important bias could be introduced to the final estimations 
  of the number of muons in measured EAS. Here, we are confident, however, that 
  the modeling is reliable at least for particles other than muons. One of the 
  reasons is that model independent studies performed in \cite{KG} have shown the 
  absence of systematic deviations between separate estimations of $N_{ch}$ for 
  vertical EAS (where the contribution of muons is not dominant) with the KASCADE 
  and the Grande arrays, although they were obtained based on independent LECF's. 
  And two further reasons are that, as we will see at the end of this section, 
  the measured $\Lambda_{ch}$ parameter shows a better agreement with the MC 
  predictions and the attenuation length for shower electrons obtained with 
  Grande data seems to be in pretty good agreement with the one derived from  
  KASCADE measurements. Hence, the problem of the observed anomalies could rely in 
  the MC estimations of the energy deposits of the muons in the KASCADE penetrating 
  detectors at different radial distances to the EAS core and distinct zenith angles. 
  If so, then a lower/higher contribution per muon to the LECF of muons would be 
  required at small/large zenith angles in order to reduce the magnitude of the 
  measured $\Lambda_\mu$ and $\alpha_\mu$ parameters and to bring the data into 
  agreement with the corresponding MC predictions. As a matter of fact, this 
  possibility is not in conflict with the general conclusion drawn at the 
  beginning of this section. 
  
   At the moment, for the following discussions, we will assume that the role of 
  the muon LECF in the deviations is small as expected from the MC simulations
  and within this context we will explore some scenarios implied by the observed 
  deviations. 
  
   \paragraph{Possible interpretations of the observed anomaly}
   One of the consequences of the mismatch between the observed and predicted 
  $\Lambda_\mu$ is that the measured muon shower size spectrum of cosmic rays attenuates 
  more slowly with increasing atmospheric depth than the simulated spectra. This
  result could be interpreted in terms of an incorrect prediction of the muon
  content of vertical and inclined EAS by the high-energy hadronic interaction 
  models. For example, $N_\mu$ could be too low for inclined showers in MC 
  simulations, or too large in case of vertical EAS.
         
     There are several possible ways to modify the muon number of EAS in 
  simulations in order to obtain a larger muon attenuation length. Some tests 
  carried out with EPOS-LHC and QGSJET-II-04 seem to indicate that 
  at KASCADE-Grande, for EAS below $\theta = 40^\circ$, we are very close
  to the region of the maximum of the muon longitudinal profile. This implies 
  that if the shower maximum is closer to the ground then $\Lambda_\mu$, as 
  reconstructed with equation (\ref{eq3}), will raise 
  and even more will become more sensitive to the position of the shower maximum. That 
  is a  geometric effect that should hold for any hadronic interaction model (at 
  least it was confirmed for EPOS-LHC and QGSJET-II-04 using EAS generated by 
  light primaries). This way, under this situation, one way to increment 
  the value of  $\Lambda_\mu$ is by increasing the interaction depth 
  of primary particles, because in this case the shower maximum would be even 
  closer to the observation level \cite{Cazon2}. A similar effect can be obtained 
  by having air showers that penetrate deeper into the atmosphere \cite{KGHadron2}. 
  The need for more penetrating air showers in simulations is a plausible 
  situation, which seems to be supported by both the analysis of the muon 
  production heights measured with the muon tracking detector (MTD) of the KASCADE 
  observatory \cite{MTD2, MTD4} and the study of the flatness of the $\bar{\rho}_\mu(r)$ 
  distributions measured with the KASCADE muon array (see \ref{appC5}). The 
  former has revealed that the maxima of the muon production height distributions
  occur at lower altitudes than in MC simulations, while the latter has shown
  that the measured muon LDF's are steeper than the ones obtained from the
  MC models. That $\Lambda_\mu$ increases when the shower maximum is closer to 
  the detector level might be verified at the KASCADE-Grande data from the 
  studies performed in \ref{appC7}. There, the variation of the muon attenuation 
  length with the atmospheric ground pressure or, equivalently, the atmospheric 
  depth was calculated. In particular, an increment of $\Lambda_\mu$ of 
  $\sim 16\, \%$ seems to be observed in the KASCADE-Grande data when decreasing 
  the ground pressure  by $\sim 8 \, \mbox{g}/\mbox{cm}^2$. Again, we should remark 
  that this only works when the maximum of the muon longitudinal profile is close 
  to the ground, which seems to be the case for the EAS measured at KASCADE-Grande.
  
    Larger $\Lambda_\mu$ values can also be achieved in simulations by requiring 
  a harder energy spectrum for shower muons at production site \cite{Pierog2013icrc}.
  It is worth to notice that if muons have a harder spectrum and hence a larger
  attenuation length, then the maximum of the muon longitudinal profile  will be 
  closer to the ground. This will further increase the magnitude of $\Lambda_\mu$ if 
  the maximum is already close to the observation level. Therefore, one of the  factors 
  which could have a remarkable effect on $\Lambda_\mu$ is
  the muon energy spectrum at production site. Amongst the models analyzed in this 
  work, QGSJET-II-04 and EPOS-LHC are the ones with the hardest spectra of muons, 
  respectively. This might be the reason why they predict the largest muon 
  attenuation lengths in comparison with the other models. There are two possible 
  ways to achieve a harder muon spectrum in MC simulations: by an increase in the 
  amount of high energy muons in the EAS  or by a decrease in the number of 
  low energy muons in the shower\footnote{In both cases the discrepancy would depend 
  also on the atmospheric grammage decreasing at altitudes closer to the height where 
  the maximum number of shower muons is reached.}. In order to 
  discriminate between these physical situations in the present models an analysis 
  of the muon data at different energy thresholds is compelling\footnote{Fortunately, 
  such analysis can be performed at KASCADE-Grande  using the surface muon array, the 
  underground muon tracking detector (MTD) and/or the tracking chambers from the 
  central detector \cite{KASCADE}. Since such analysis is underway, further hints to 
  check the deficiencies of the models concerning the energy spectrum of muons 
  may be obtained in the future.}.

    In addition to the muon attenuation length, the $\alpha_{\mu}(r)$
  coefficients may also provide some information about  overall 
  differences between the energy spectrum of muons from MC and measured
  data. What we have seen in fig.~\ref{Results_absorption_New} is a 
  deviation, which seems to increase with the radial distance $r$ to the
  shower core (measured at the shower plane). This behavior might point out 
  important deficiencies of the hadronic interaction models in describing 
  also the correct proportion of low energy muons to high energy ones  
  but as a function of the lateral distance, $r$. At closer distances to the 
  EAS core, fig.~\ref{Results_absorption_New} seems to suggest that an 
  increase in the amount of high energy muons could be appropriate at least 
  for QGSJET-II-04 in order to reproduce the experimental data on 
  $\alpha_{\mu}(r)$, since the contribution of high energy muons to the 
  LDF's becomes more important close to the shower axis \cite{Meurer, MTD2005, 
  Drescher03}. 

   On the other hand, at larger distances from the EAS core, where low energy 
  muons are more important, the aforesaid figure seems to indicate that
  modifications are necessary for all the studied models. In this case, 
  the observed deviations might call not only for a reduction in the amount 
  of low energy muons in the simulated EAS, but also for an increment in 
  the content of muons at higher energies.  The latter in view of the fact 
  that as the zenith-angle increases, both the experimental energy threshold 
  and the mean energy of the muons rise \cite{Cazon2}. This way, the muon 
  content in inclined showers becomes more sensitive to the high energy 
  part of the spectrum, which can lead to a rise in the value of
  $\alpha_{\mu}(r)$  at large distances from the core if the number 
  of high energy muons is increased.

 \paragraph{Role of the low-energy hadronic interaction models}
  We are assigning the discrepancy between the measurements and the simulations 
  to the influence of the high-energy hadronic interaction models. But, as we 
  measure muons with a $230 \, \mbox{MeV}$ energy threshold at sea level, both
  the muon number of EAS  and the lateral density of muons are affected by 
  the decay products of low energy charged mesons from the last part of the 
  shower development \cite{MTD2005, Engel01, Haungs01}. Thus, a change in the 
  description of the low-energy hadronic interactions might also have important 
  modifications to the magnitudes of $\alpha_\mu(r)$ and $\Lambda_{\mu}$, 
  mainly at large distances from the core. Therefore, low-energy hadronic interaction 
  models might be playing a relevant role in the discrepancy. The issue will be 
  investigated in detail in forthcoming studies.

  \paragraph{Consequences of the $\Lambda_\mu$ anomaly}
  Due to the rapid attenuation of the simulated data in comparison 
  with the actual one, the discrepancy has some implications for the energy 
  spectrum and the composition studies of cosmic rays when air-shower data from 
  different zenith angles are employed. In the first case, the anomaly will 
  introduce a shift to higher energies on the primary spectra of cosmic rays
  reconstructed with $N_\mu$ data from inclined showers.
  This shift was observed in the analysis of \cite{KGNchNmu}, where it was
  shown that, for measured EAS  with $\theta < 40^\circ$, the anomaly introduces
  an uncertainty of $6.5 \, \%$ at $10^{16} \, \mbox{eV}$ and $10.9 \, \%$
  at $10^{17} \, \mbox{eV}$ in the respective all-particle cosmic ray flux
  when using QGSJET-II-02 as a framework for the energy calibration of the
  data.
   
    As a consequence of the above shift, the elemental composition of
  cosmic rays as inferred from the measured data using the high-energy hadronic
  interaction models appears heavier with increasing zenith angles. Indeed,
  inside the framework of the discussed hadronic interaction models, the analyses
  of the muon densities at different $N_{ch}^{CIC}$ bins and zenith-angle intervals
  (c.f. fig.~\ref{DensitiesKG_vs_Models_zenith}) show that the actual $\bar{\rho}_\mu(r)$ 
  distributions move gradually towards a heavier composition for inclined showers. 
  As an example, EPOS-LHC favors a light composition at around $10^{17} \, \mbox{eV}$ 
  for vertical EAS, while for inclined showers the model indicates that a mixed 
  composition is dominant in the experimental data at roughly the same energy. 

  The source of disagreement between the measured and the predicted $\Lambda_\mu$ 
  in KASCADE-Grande could be also responsible for another anomaly detected at higher 
  energies by the Pierre Auger collaboration. Measurements performed with the Auger 
  observatory have shown an excess of the total $\mu$-content ($E_\mu > 0.3 \, 
  \mbox{GeV}$) in experimental data at ultra-high energies in comparison with 
  expectations from modern MC simulations. Such anomaly has been observed also
  with the Yakutsk array ($E_\mu > 1 \, \mbox{GeV}$) \cite{Yakutsk}. The 
  discrepancy seems to be energy \cite{AugerMuonExcess2} and zenith-angle dependent 
  \cite{AugerMuonExcess} and can not be described by any of the available hadronic 
  interaction models. Remarkably the largest deviations observed with the Auger 
  detector between MC predictions and experimental data  seems to occur for inclined 
  showers and the highest energies. The latter might imply that model predictions 
  can not even match the muon attenuation length of EAS at ultra-high energies and 
  that such effect could evolve with the shower energy. A possible energy dependence 
  of the $\Lambda_\mu$ anomaly will be investigated in future studies at 
  KASCADE-Grande by adding EAS data with shower energies below $10^{16}\, \mbox{eV}$ 
  from the KASCADE array.

   With the aim of having a better understanding of the muon deviation measured at 
  the KASCADE-Grande detector, independent studies from other observatories on the 
  matter could be useful, specifically, at the energy range explored in this paper,
  using the current MC models. Unfortunately, such studies are absent at the 
  moment. Muon data exist around $E = 10^{17}\, \mbox{eV}$  from HiRes-MIA 
  ($E_\mu > 850 \, \mbox{MeV}$) \cite{MIA}, the EAS-MSU array ($E_\mu > 10 \, 
  \mbox{GeV}$) \cite{EAS-MSU} and the IceTop  ($E_\mu > 200 \, \mbox{MeV}$) experiment
  \cite{ICETOP}, but the analyses have been restricted only to look for a possible 
  muon excess in the measured data over model predictions in a zenith-angle 
  independent way. Hence, it is not possible to say whether the  $\Lambda_\mu$
  anomalies are also present at the experimental conditions (i.e., muon energy 
  thresholds, radial ranges and air grammages) of such observatories. Undoubtedly 
  these information  would help to provide a wider picture of the above problem
  and narrow down the number of possible solutions.

  \paragraph{Remarks about $\Lambda_{ch}$}
   Regarding our results corresponding to the attenuation length of $N_{ch}$ 
  (see \ref{Lambdach}), we see a better agreement between the experiment and 
  the MC simulations than in the case of $\Lambda_\mu$. In fact, the 
  deviations of the measured $\Lambda_{ch}$ from model predictions are 
  less than $+1.39 \, \sigma$. By comparing the results obtained with 
  the QGSJET-II models, we observe that the post-LHC improvements 
  performed in the last version of QGSJET-II did not spoil the agreement 
  between the predicted and measured values of $\Lambda_{ch}$. That is an 
  important constraint that, among other ones (such as the electron-muon 
  correlations \cite{KGHadron}) must be supervised  when applying 
  modifications to the models. 

   Since, at the energies and zenith angles involved in the analysis,
  $N_{ch}$ is dominated by shower electrons, the fact that the value
  of $\Lambda_{ch}$ is closer to the predictions of the models
  might indicate that the  cause of the anomaly observed in 
  the longitudinal development of $N_\mu$ in the atmosphere has
  not a strong impact on the atmospheric attenuation of the 
  electromagnetic component of the EAS.
  
   We observed that the magnitude of $\Lambda_{ch}$ is smaller 
	than $\Lambda_\mu$. This is expected due to the stronger attenuation of $N_e$
  in comparison with $N_\mu$ and the dominance of electrons over muons in
  $N_{ch}$ for our selected data set. Following the same reasoning, we
  should also expect  $\Lambda_{ch}$ to be closer to  the attenuation 
  length for the number of electrons, $\Lambda_e$. In order to verify
  the consistency of the results, we calculated $\Lambda_e$ and
  compared it with $\Lambda_{ch}$. By applying the CIC method to the 
  experimental data on $N_e$, we obtained  $\Lambda_e = 192 \pm 8 \, 
  \mbox{g}/\mbox{cm}^2$ from fits to the data  in the interval 
  $\log_{10} N_e = [5.9, 7.1]$ (only the error from the global fit is 
  quoted)\footnote{The result is in full agreement with the measurements 
  performed with KASCADE at lower energies. In this case, $\Lambda_e$ was found
  to vary between  $170$ and $192 \, \mbox{g}/\mbox{cm}^2$ using the 
  CIC method in the interval $4.5-6.5$ of $\log_{10}(N_e)$ \cite{LambdaNe}.}.
  This value is just $1.1 \, \sigma$ below $\Lambda_{ch}$. Therefore, in
  light of the  previous discussions, we found that, inside the 
  corresponding  experimental uncertainties, the measurements of 
  $\Lambda_{ch}$ and $\Lambda_e$ are not inconsistent between each other.

  \section{Implications for the features of hadronic interaction models}
  \label{Consequences}

   The physical origin of the $\Lambda_\mu$ discrepancy is not yet clear. Insofar, 
  as the attenuation of muons in matter is concerned, this process is almost 
  completely described by QED (with the exception of deep inelastic scattering, 
  which contributes to the energy loss only less than $1$\, \%). Assuming that
  electromagnetic processes in air showers are well described by the EGS4 \cite{EGS4} 
  code used in COSIKA, any inconsistency between the measured and predicted muon 
  attenuation lengths must be attributed to the modeling of hadronic interactions 
  or to the description of the hadronic shower development in the atmosphere. This 
  way, our results would indicate that the high-energy hadronic interaction models 
  QGSJET-II-02, SIBYLL 2.1, EPOS-LHC and QGSJET-II-04 need modifications to 
  resolve the discrepancy with the muon data from KASCADE-Grande. 
	  
    In the last section  we  discussed some possible modifications of EAS 
  characteristics in the models, which might help to solve the muon attenuation 
  length problem observed at KASCADE-Grande, e.g., an increase in the depth of the 
  first hadronic interaction in the EAS, a deeper muon production height and a 
  harder muon energy spectrum at production site. Now, we will discuss some changes 
  of the characteristics of the internal parameters of the high-energy hadronic 
  interaction models that might produce the variations in the EAS observables 
  desired to explain the $\Lambda_\mu$ anomaly.
   
    In order to change the depth of the first interaction of the incident 
  cosmic ray, $X_1$, the relevant parameter is the cross section for inelastic 
  collisions with air, $\sigma_I$. Since, $X_1 \propto 1/\sigma_I$ \cite{Grieder}, 
  the depth of the first interaction can be increased by reducing  $\sigma_I$.  
  However, in this regard, there is not much room left due to the strong 
  constraints set on the models by the LHC proton-proton data \cite{Totem, Atlas}. 
  Consequently, this possibility might just have a minor contribution to the 
  discrepancy after all. 
   
   A bigger effect could be obtained from a deeper muon production depth (MPD) 
  in the atmosphere, $X^{\mu}$. The latter can be achieved by modifying the 
  description of pion-nucleus interactions, which is an important source of 
  uncertainty in the models. More specifically, from detailed studies performed in 
  \cite{PierogICRC15, Ostap2016}, $X^{\mu}$ can be augmented principally through 
  an increase of pion elasticity, a smaller pion-air inelastic cross section, 
  harder secondary hadron spectra in pion-air collisions and/or a copious 
  production of (anti-)baryons. The last option, however, it is not useful to
  enlarge $\Lambda_{\mu}$ as we will explain later, therefore it might
  be discarded as a possibility to reduce the anomaly.  The remaining options,
  on the other hand, could be coherent with an increase of  $\Lambda_{\mu}$. 
  Here, special care must be taken to be consistent also with the Pierre 
  Auger measurements on the average value of $X^{\mu}_{max}$, i.e. the 
  maximum of the $X^{\mu}$ profile  \cite{AugerXmaxMuon1}. In case of EPOS-LHC, 
  for example, a further increase of $X^{\mu}$ is not supported by the Auger 
  data. The reason is that the respective model predictions are well above the 
  experimental values at ultra-high energies. In 
  this case a  reduction of $X^{\mu}$ is  imperative. This can be achieved,
  for example, through a decrease of the elasticity in pion interactions
  \cite{PierogICRC15} and/or a suppression of forward production of 
  baryon-antibaryon pairs \cite{Ostap2016}. The first change could lead to an 
  opposite effect in $\Lambda_\mu$ to the one desired, while the second one 
  could be coherent with the intended objective.

    Of great importance for the problem could be the hadron and resonant production 
  processes that keep energy of the shower in the hadronic channel  and which could 
  be misrepresented in the models. They can modify the expected energy  spectra of 
  muons and, hence, the predicted muon attenuation lengths. To this category belongs 
  the creation of (anti-)baryons in pion-air interactions. It is known that the 
  abundant production of baryon-antibaryon pairs enhances $N_\mu$ \cite{epos199, 
  Allen13}, but it also increases the proportion of low energy muons in the shower. 
  Thus, if it is overestimated, it might shorten the muon attenuation length and, 
  hence, it could increase the $\Lambda_\mu$ discrepancy.  That seems to be happening 
  in EPOS-LHC as it is suggested by Auger data on $X^{\mu}_{max}$. In principle,
  solving the problem of low energy muons in EPOS-LHC will put  $X^{\mu}_{max}$
  higher in the atmosphere in agreement with the Auger observations, but it 
  will also produce a harder muon energy spectrum and hence an increase of the 
  distance between the MPD (where the muons are created) and the maximum 
  of the muon longitudinal profile putting the latter closer to the ground, 
  which is an important factor to increase $\Lambda_\mu$.

     A further mechanism that changes the muon energy spectra of EAS and is not 
  well described in some models is the production of $\rho^0$ resonances in 
  pion-nucleus interactions. This process could also prove to be valuable to reduce 
  the proportion of low energy muons at ground and to increase the magnitude 
  of $\Lambda_\mu$ in the models.  The reason is that this mechanism enhances 
  the production of high energy muons during the early stages of the EAS. After
  production, the $\rho^0$ mesons decay almost immediately into a pair of charged 
  pions \cite{Allen13}. At the early stages of shower development, these pions 
  have a bigger probability to decay than to interact in the air (because the 
  density of the atmosphere is low at high altitudes) resulting in the creation of 
  high energy muons \cite{Ulrich11}.  In particular,  QGSJET-II-02, SIBYLL 2.1 and 
  EPOS-LHC underestimate the fixed-target experimental results  on the very 
  forward spectrum of $\rho^0$-mesons in pion-nucleus interactions  \cite{NA61}.  
  Consequently, an enhancement of the above mechanism in these high-energy hadronic 
  interaction models is necessary. This improvement might decrease the $\Lambda_\mu$
  differences between models and experiment in these cases.  

   The transverse momentum ($p_t$) distributions of charged pions generated in 
  pion-nucleus collisions may also need further tuning inside the current 
  high-energy hadronic interaction models, as revealed by the results of 
  the NA61/SHINE experiment about the spectra of charged pions in $\pi^{-}-C$ 
  interactions \cite{NA61}. The $p_t$ distributions of $\pi^{\pm}$'s 
  have a relevant influence on the muon LDF's. Hence, it seems  plausible that 
  they would have also some impact on the magnitude of $\Lambda_\mu$ as extracted 
  from the local measurements of muons in EAS at KASCADE-Grande.

   Finally, one could question the role of the approximations implemented in 
  EGS4 \cite{EGS4} in the $\Lambda_\mu$ discrepancy. This is an open issue, which 
  has not been fully investigated. One might argue, therefore, that
  the observed anomaly could receive some contributions from an inaccurate 
  description of the electromagnetic process behind both the attenuation of 
  muons in the atmosphere or the photoproduction of low energy muon pairs. In 
  spite of that, we might stress the role of the hadronic interaction models in 
  the observed anomaly, as there are no direct 
  experimental evidence for the existence of problems with such approximations
  which could give further support to the aforesaid hypothesis.

   \section{Conclusions}
  \label{Conclusions}

   In this paper, the QGSJET-II-02, SIBYLL 2.1, EPOS-LHC and QGSJET-II-04 
  high-energy hadronic interaction models have been tested by comparing their 
  predictions for the attenuation length of muons in EAS with the measurements 
  performed with the KASCADE-Grande experiment at the energy interval
  $E \approx 10^{16.3} - 10^{17.0} \, \mbox{eV}$. In particular, it was found 
  that the experimental $\Lambda_\mu$ value is above $+2.04 \, \sigma$  and  
  $+1.99 \, \sigma$ from the QGSJET-II-02 and SIBYLL 2.1 expectations, 
  respectively, and just   $+1.48 \, \sigma$  and  $+1.34 \, \sigma$ from the 
  corresponding QGSJET-II-04 and EPOS-LHC predictions. The above implies that the 
  studied pre-LHC models do not match the measured value of $\Lambda_\mu$, 
  while the post-LHC models are in relatively good agreement with the data.
  Despite of the latter, however, the fact that the expected muon attenuation 
  lengths from the post-LHC models are below the actual value seems to 
  suggest that these models need further tuning to describe the 
  KASCADE-Grande data. 
  
   To investigate the possible origin of the above deviations, predictions for 
  the average muon densities at different zenith angles and $E \approx 10^{16.9} - 
  10^{17.2} \, \mbox{eV}$ along attenuation curves in shower size were also 
  confronted with the experiment. In general, it was found 
  that the measured absorption lengths of the  aforesaid mean muon density 
  distributions become bigger than the predictions of the high-energy hadronic 
  interaction models analysed in this work at large distances from the EAS core. 
  According to complementary tests performed with MC simulations, we found that 
  the aforesaid discrepancies could be the cause of the observed differences 
  between the measured and the expected $\Lambda_\mu$ values.
    
  Finally, the attenuation length of $N_{ch}$ was also measured and compared
  with the predictions of the hadronic interaction models. In this case,
  good agreement between the experiment and expectations was observed
  with differences ranging from $+0.51 \, \sigma$ to $+1.39 \, \sigma$.
  
  In conclusion, the QGSJET-II-02, SIBYLL 2.1, EPOS-LHC and 
  QGSJET-II-04  hadronic interaction models do not reproduce consistently the 
  zenith-angle behavior of the selected  KASCADE-Grande data on the local muon 
  content (with threshold energies $E_\mu \geq 230 \, \mbox{MeV}$ at vertical 
  incidence) of EAS.

 \section*{Acknowledgments}

  The authors would like to thank the members of the engineering and technical 
 staff of the KASCADE-Grande Collaboration, who contributed to the success of the
 experiment. The KASCADE-Grande experiment was supported in Germany by the BMBF 
 and by the ’Helmholtz Alliance for Astroparticle Physics - HAP’ funded by the 
 Initiative and Networking Fund of the Helmholtz Association, by the MIUR and 
 INAF of Italy, the Polish Ministry of Science and Higher Education, the 
 Romanian Authority for Scientific Research UEFISCDI (PNII-IDEI grants 271/2011 
 and 17/2011), and the German-Mexican bilateral collaboration grants (DAAD-CONACYT 
 2009-2012, 2015-2016). J.C.A.V. acknowledges the partial support of CONACyT 
 (grant CB-2008/106717) and the Coordinaci\'on de la Investigaci\'on Cient\'\i fica 
 de la Universidad Michoacana.

 \appendix

  \renewcommand{\thefigure}{A\arabic{figure}}
  \setcounter{figure}{0}
  
  \section{Muon Correction function}
  \label{appNmuCF}
 
    The location of the muon detectors at the fringe of the Grande array, the limited 
  size of the muon array and the detection and reconstruction procedures introduce a 
  systematic error on the muon size, which depends on the arrival angle, the core 
  position and the shower size. In order to improve the accuracy of the EAS observable 
  and eliminate, as much as possible, the influence of the muon systematic errors on 
  the study, a muon correction function is applied. The correction is achieved 
  by using a single function that is derived from MC data, in particular, the 
  QGSJET-II-02 data set, which has a better statistics and hence a reduced statistical 
  error. Herein the shape of the function is parameterized in terms of the shower 
  core position at ground, the shower size and the EAS zenith and azimuth angles. 
  In the derivation of the correction function, the mixed composition scenario is assumed 
  obeying to the uncertainty of the elemental abundances in cosmic rays. Also a spectral 
  index $\gamma = -3$ is employed.

 \begin{figure}[!t]
 \centering
 \includegraphics[width=3.2in]{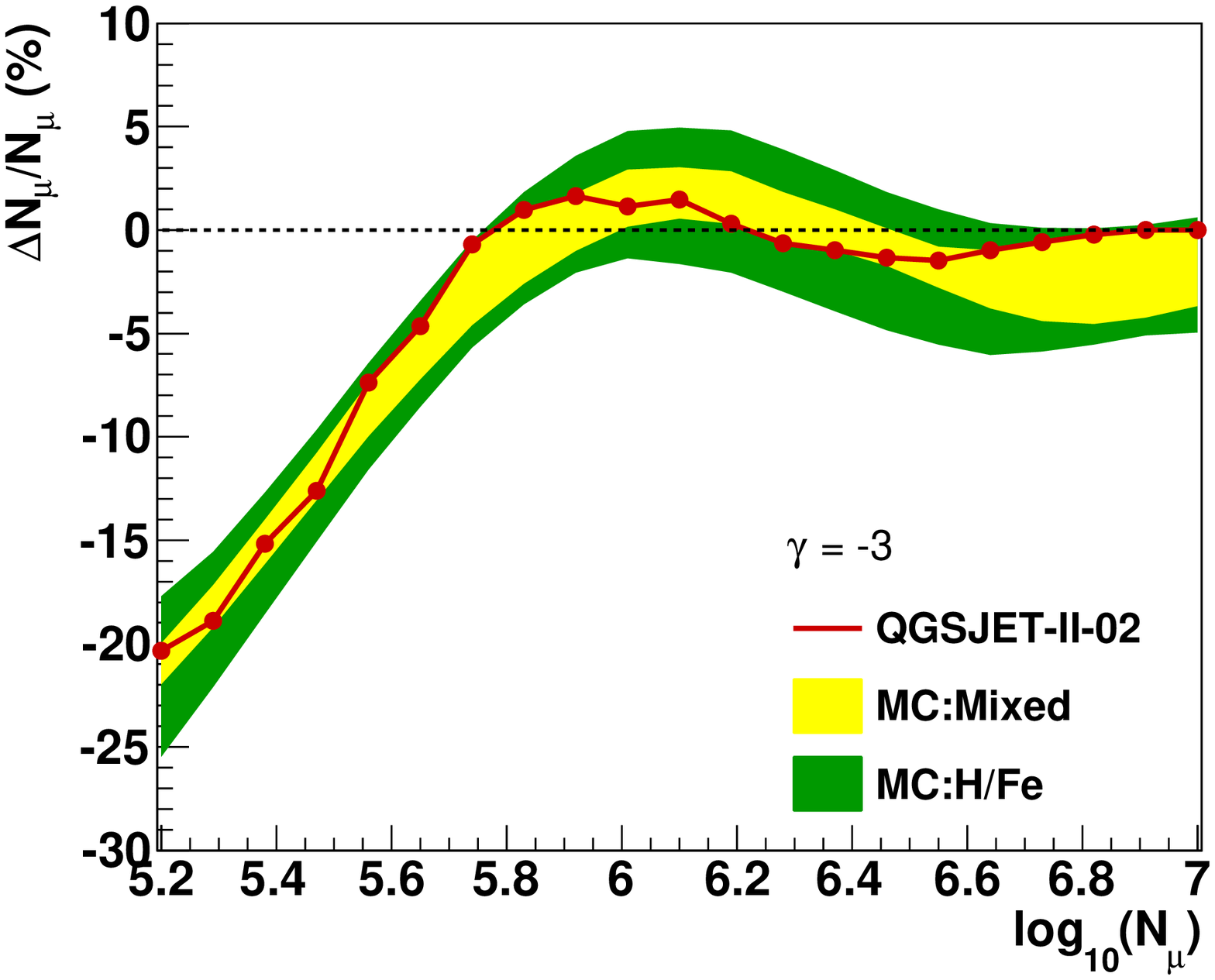}
 \includegraphics[width=3.2in]{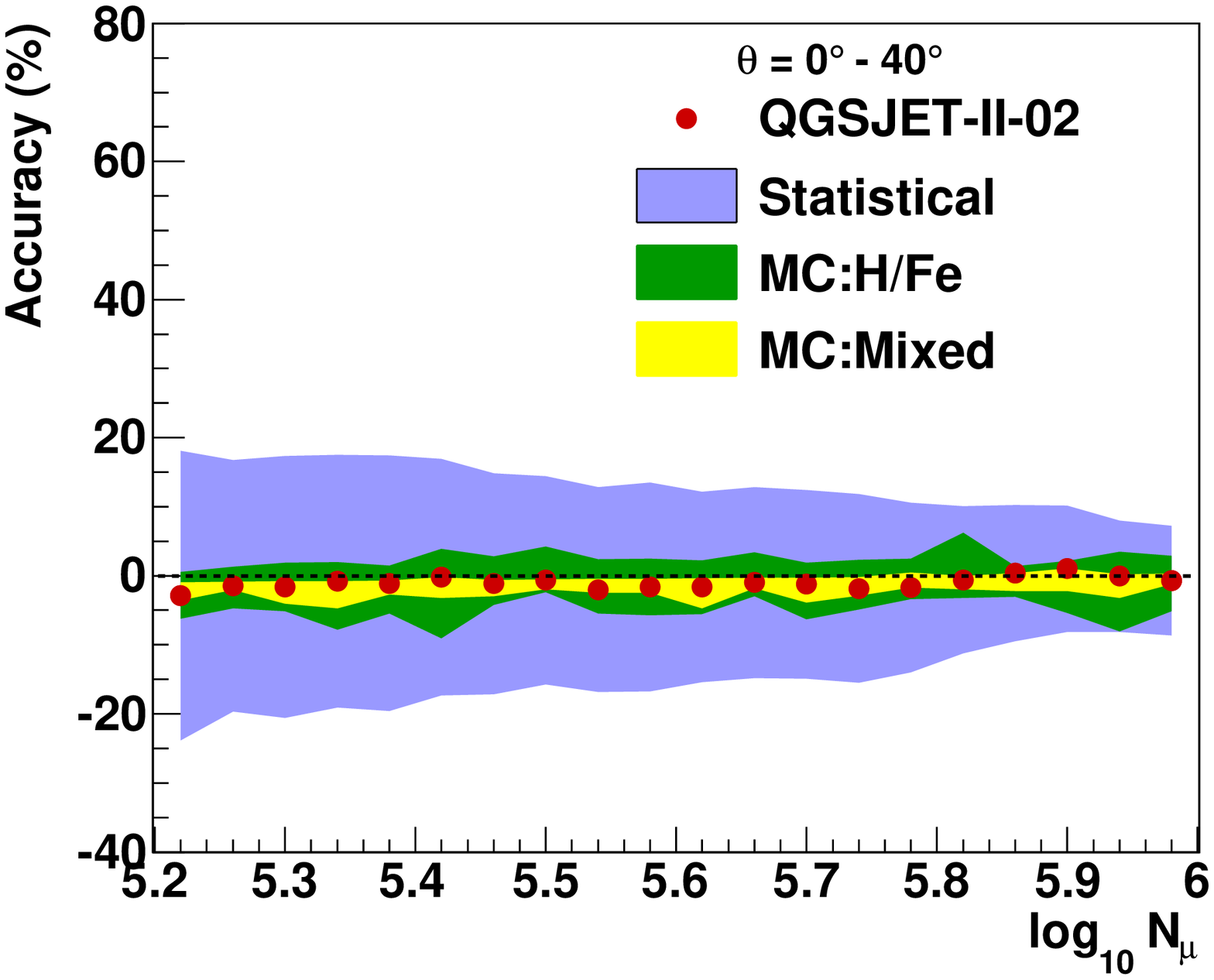}
  \caption{Left: Mean value of the muon correction function against the uncorrected
  $N_\mu$ for different hadronic interaction models assuming a primary spectral 
  index $\gamma = -3$. The function was evaluated for the KASCADE-Grande fiducial 
  area and for a solid angle with $\theta = [0^{\circ}, 40^{\circ}]$. Right: 
  Mean value of the systematic errors for the corrected muon number plotted 
  as a function of the corrected $N_\mu$. In both figures, the points represent the
  results for QGSJET-II-02 assuming mixed composition. The error band labeled as 
   \textit{mixed} covers the range of variation of the results
  when a mixed composition scenario is assumed and the different hadronic 
  interaction models studied in this paper are individually employed: QGSJET-II-02, 
  SIBYLL 2.1, EPOS-LHC and QGSJET-II-04. On the other hand, the error band labeled as 
  \textit{H/Fe} covers the expectations for pure hydrogen and iron nuclei. Finally, 
  the error band labeled as \textit{statistical} that appears on the right figure 
  is the statistical error band for the 
  results of QGSJET-II-02 shown with points.}
 \label{MuonCF1}
\end{figure}

 \begin{figure}[!t]
 \centering
 \includegraphics[width=3.2in]{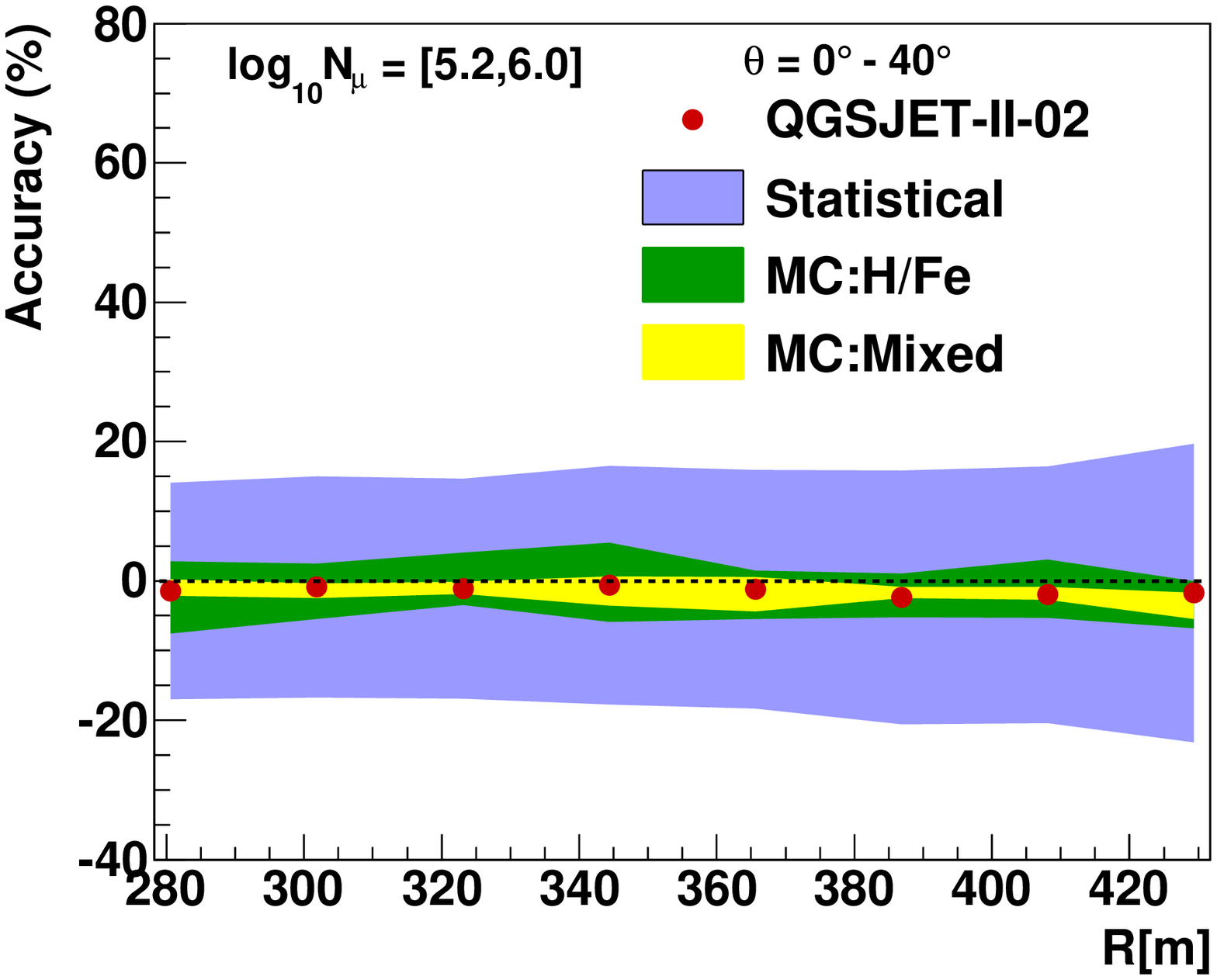}
 \includegraphics[width=3.2in]{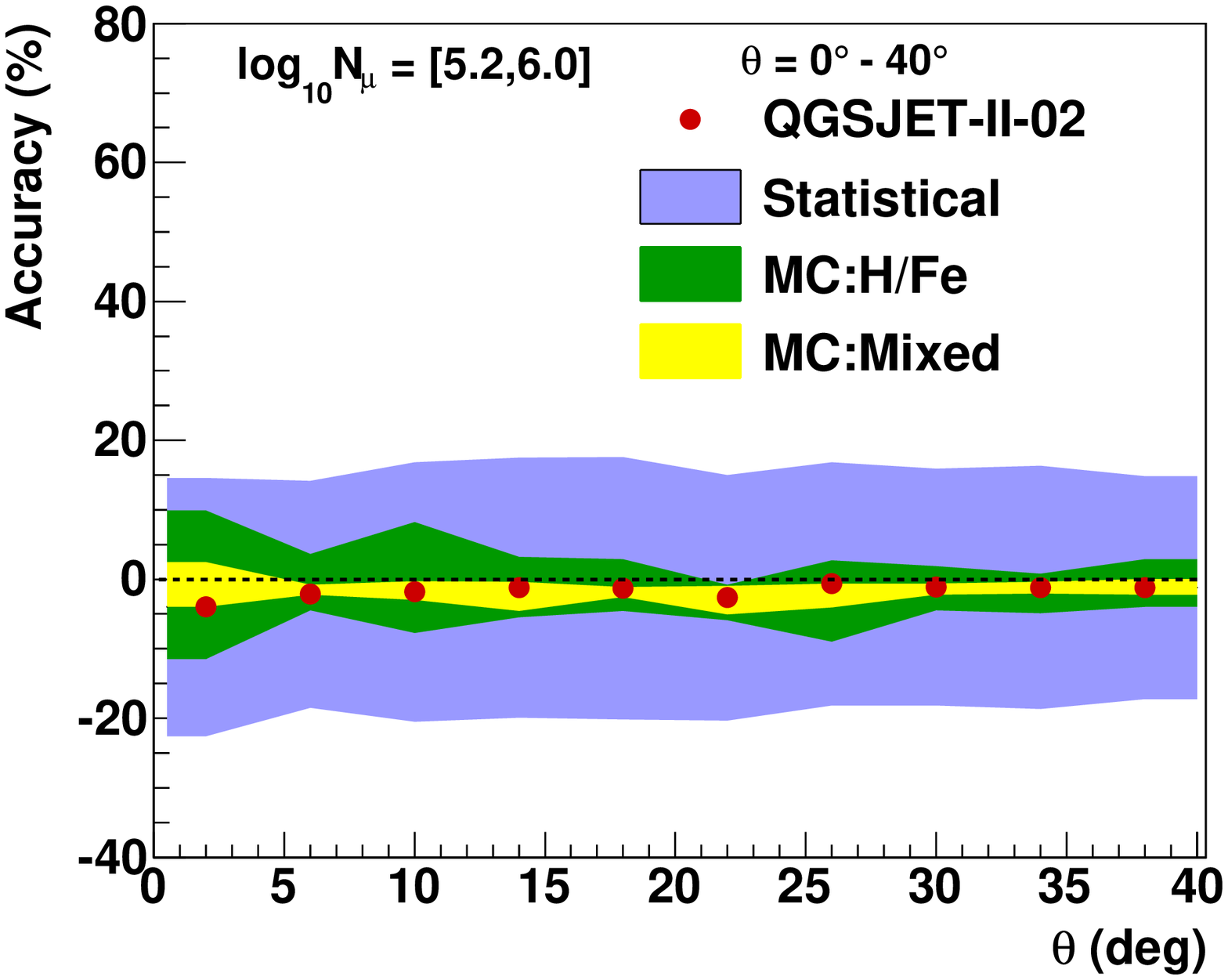}
  \caption{Mean value of the systematic uncertainties for the corrected muon 
  number expected for the fiducial area of KASCADE-Grande and the zenith-angle 
  interval $\theta = [0^{\circ}, 40^{\circ}]$. Errors are presented for corrected
  muon numbers within the interval $\log_{10} N_\mu = [5.2, 6.0]$, where the 
  analysis of $\Lambda_\mu$ was performed. The systematic errors are
  shown as a function of the core distance to the KASCADE center (left) and
  the shower zenith angle (right). In both figures, the points represent the
  results for QGSJET-II-02 assuming mixed composition. The meaning of the error 
  bands is the same as in fig.~\ref{MuonCF1}.}
 \label{MuonCF2}
\end{figure}

   The use of a single correction function on the muon data is justified since it 
  is nearly independent of the composition and the hadronic interaction models 
  explored here. Using other hadronic models and/or different composition assumptions 
  just introduces small relative differences (within $\approx \pm 5\%$) in the correction 
  function. This can be appreciated 
  in fig.~\ref{MuonCF1} (left), where the mean value of the muon correction function 
  from QGSJET-II-02 is plotted against the uncorrected $N_\mu$ for showers with cores 
  inside the KASCADE-Grande fiducial area and EAS axes between $\theta = 0^{\circ}$ and 
  $40^{\circ}$. The plots are shown along with two error bands that cover the range 
  of results for alternative correction functions derived individually from different 
  hadronic interaction models and composition scenarios. In fig.~\ref{MuonCF1},
   (left) with low $N_\mu$, the correction on the reconstructed
  muon number is large. That is because for low energy events located 
  outside the KASCADE detector area the number of muons is overestimated. The reason 
  is well known and it is due to the fact that the LDF that is used to get $N_\mu$ 
  on an event-by-event basis is steeper than the expected distribution of local 
  muon densities  for the EAS with the abovementioned characteristics \cite{KG}. 
  At high energies, this difference decreases, which reduce the uncertainty of the 
  reconstructed $N_\mu$ and thus the magnitude of the applied muon correction as
  observed in fig.~\ref{MuonCF1} (left).  

   The mean systematic errors of the corrected muon number are displayed in 
  fig.~\ref{MuonCF1} (right) and  fig.~\ref{MuonCF2} as a function of the the muon 
  size, the core position and the shower zenith angle in the full efficiency 
  and maximum statistics regime. We can see that the final systematic errors are 
  less than $10\%$.  Although, this remaining bias is small we have not neglected it
  and have considered it in the evaluation of the final uncertainties of the
  muon attenuation length.\\

 \section{Error estimation on $\Lambda_\mu$}
 \label{appA}
 
  \renewcommand{\thefigure}{B\arabic{figure}}
  \setcounter{figure}{0}
  
 \setcounter{table}{0}
 \renewcommand{\thetable}{B\arabic{table}}

  In table~\ref{tabAppB} the total uncertainties for $\Lambda_\mu$ 
  are shown with the individual contributions from statistical
  and systematic errors. For the case of MC simulations 
  the total errors vary in the range from $\approx -30 \, \%$
  to $\approx +28 \, \%$, while for experimental data they are
  found to be between $-20 \, \%$ and $+19 \, \%$. In the following, 
  we will list both the main statistical and systematic uncertainties 
  that we have taken into account in the above estimations and briefly 
  describe how they were calculated.

   	\textit{Statistical error}. For the estimation of the influence of 
  statistical fluctuations on
  the measured $\Lambda_\mu$, $\Phi_\mu$ intensities are randomly drawn from 
  the original KASCADE-Grande muon shower size spectra by allowing the number 
  of events per $N_\mu$ interval and angular bin to fluctuate according to a 
  Poisson distribution. For each trial, the integral intensities are then 
  calculated from the drawn $\Phi_\mu$ spectra for each zenith-angle 
  interval according to eq.~\ref{eq2}. Afterwards, the attenuation 
  length is estimated with the usual method. The statistical error is 
  therefore computed from the observed variability of $\Lambda_\mu$ 
  after 50 trials. In case of MC simulations, the procedure is similar,
  but with a single difference: as the MC data are weighted we use
  the formalism of the equivalent number of unweighted events \cite{Zech}
  in the construction of the trial spectra, which allow us 
  to properly evaluate the influence of statistical uncertainties on 
  the expected $\Lambda_\mu$ values. 
  
   Let $N$ be the number of simulated events in a given $N_\mu$-bin   
  and $w_j$, the individual weights of such events, where $j = 1, ..., 
  N$. Then the number of events in the corresponding bin of the 
  weighted histogram is $N^{ev} = \sum^{N}_{j = 1} w_j$, with 
  $\sigma(N^{ev}) = \left[ \sum^{N}_{j = 1} w^2_j  \right]^{1/2}$, the respective statistical 
  error. In general, $N^{ev}$ does no follow a Poisson distribution, 
  therefore, we replaced it by the equivalent number of unweighted 
  events $\tilde{N}^{ev} = (N^{ev})^{2}/\left[ \sigma(N^{ev}) \right]^2$. 
  This quantity is Poisson distributed and has the same relative 
  statistical uncertainty as $N^{ev}$. From here, we obtain the trial 
  $\Phi_\mu$ spectra that we require by allowing $\tilde{N}^{ev}$ to 
  fluctuate in each $N_\mu$-bin according to a Poisson distribution and 
  after multiplying the result with a corresponding factor 
  $w_r = N^{ev}/\tilde{N}^{ev}$ to properly normalize the content 
  of the bin.

    \begin{table}[!t]
    \begin{center}
    \caption{Systematic and statistical uncertainties on the
     predicted and experimental muon attenuation lengths.     
     Contributions of the systematic errors to the total uncertainty are 
     listed individually.}
   \scriptsize
   \begin{tabular}{l|c|c|c|c|c}
   \hline
    &  QGSJET-II-02 & QGSJET-II-04 & SIBYLL 2.1 &  EPOS-LHC& KG data \\ 
   \hline
   \textbf{Statistical error ($\%$)}  &&&&&\\    
   Statistical fluctuations      &$\pm 4.29$&$\pm 8.51$&$\pm 7.51$ &$\pm 4.50$ &$\pm 6.74$\\
   &&&&&\\    
   \textbf{Systematics ($\%$)}  &&&&&\\
   Muon systematics              &$+0.04$   &$-2.30$   &$-4.78$   &$-2.53$    &$+13.55/-10.60$\\
   Core far from KASCADE         &$+2.37$   &$-0.11$   &$+2.57$   &$+5.03$    &$+11.89$\\
   ($R = [360, 440] \, \mbox{m}$)&&&&&\\
   Core close to KASCADE         &$-3.38$   &$+0.93$   &$-5.90$   &$-5.15$    &$-10.73$\\
   ($R = [270, 360] \, \mbox{m}$)&&&&&\\
   Bin size                      &$+6.14$   &$+3.70$   &$-2.05$   &$+0.29$    &$+6.79$\\
   Global fit                    &$\pm 4.96$ &$\pm 5.40$   &$\pm 5.39$   &$\pm5.05$    &$\pm5.60$\\
   Muon correction function uncertainties        
                                 &$+1.34$   &$-1.11$   &$+0.78$   &$-2.25$    &$-2.54$\\
   Broader zenith-angle interval &$+1.61$   &$-1.94$   &$-1.21$   &$+1.17$    &$-2.42$\\
   (Four angular bins)&&&&&\\
   Number CIC cuts               &$+1.12/-0.59$   &$+2.29/-0.92$   &$+0.38/-0.30$   &$+0.11/-2.06$    &$+1.40$\\
   Narrower CIC interval         &$-0.28$   &$-2.90$   &$-2.95$   &$+2.88$    &$-0.61$\\
   ($\log_{10}N_\mu \approx [5.4, 6.0]$) &&&&&\\
   Spectral index uncertainties  &$+1.24/-0.62$   &$+2.59/-0.71$   &$+1.96/-3.26$   &$-1.22$    &$-$\\
   ($\Delta \gamma = \pm 0.2$)   &&&&&\\
   Composition                   &$+10.91/-9.19$  &$+25.96/-27.57$  &$+0.07/-7.88$  &$+18.98/-10.76$   &$-$\\
   \textbf{Total ($\%$)}   &&&&&\\
                                 &$+14.57$  &$+28.32$  &$+\,9.84$   &$+21.01$    &$+19.46$\\
                                 &$-11.82$  &$-29.70$  &$-15.18$    &$-14.32$     &$-19.71$\\    
   \hline
   \end{tabular}
   \label{tabAppB}
   \end{center}
  \end{table}
  
   \textit{Error from the remaining systematic bias of the corrected muon 
   number}. Its contribution to the total error is obtained by propagating the 
  uncertainties of the corrected $N_\mu$ to the differential spectra and then 
  to the integral spectra employed in the derivation of the attenuation length. 
  The systematic biases of the corrected $N_\mu$ were estimated from MC data
  (see, for example, figs.~\ref{MuonCF1} and ~\ref{MuonCF2}). 
  In case of simulations, they were applied  in correspondence 
  with the composition scenario and the hadronic model under study. In contrast, 
  for measured data, all $N_\mu$ systematic biases that are predicted by the 
  hadronic models for several composition scenarios (i.e. five pure primary 
  nuclei, from H to Fe, and a mixed composition assumption) were used. We 
  then compared the biases introduced in the measured muon attenuation 
  length by these different hypotheses. The highest and lowest deviations 
  are quoted as the errors of the measured $\Lambda_\mu$ from the uncertainties 
  of the corrected muon number. We proceeded in this way due to the lack of 
  knowledge of the actual systematic bias of the observed $N_\mu$, the real
  hadronic interaction model and
  the primary composition of cosmic rays. As a matter of fact, this is the
  reason why the contribution of the systematic bias of the corrected 
  $N_\mu$ is one of the biggest ones to the total experimental error. For 
  MC simulations, on the other hand, this contribution was found to be 
  small. The latter due to the fact that both the composition and the model 
  are known. 
  
   \textit{Influence of the EAS core position in the systematic
  uncertainty of $\Lambda_\mu$}. The contribution of this systematic
  source was investigated by dividing the central area into two smaller 
  regions with approximately the same statistics. The division was done by 
  applying a radial cut around $360 \, \mbox{m}$ from the center of the 
  KASCADE array. To estimate the systematic errors, the muon attenuation 
  lengths from the data collected on each surface were calculated independently 
  and were later compared with the standard result for the whole area. The 
  two differences obtained in this way were then cited independently
  as the errors due to the EAS core position. Using this analysis,
  we found a dependence of the measured attenuation length on the radial 
  distance to the KASCADE center (see table~\ref{tabAppB}), which is the
  origin of a major contribution to the total experimental uncertainty. 
  By performing additional studies, we arrive at the result that the aforesaid 
  EAS core dependence is due to a small decrease of the estimated number of muons, 
  which is more important for vertical showers, as we move far away from 
  the center of the KASCADE array. In MC data, this behaviour was not observed. 
  In this case, the error analysis yielded just a mild dependence of the 
  predicted $\Lambda_\mu$ with the EAS core position.
   
   \textit{Uncertainty from the CIC method}. This contribution covers the 
  propagation of errors arising from the global fit and the variation of 
  the results with the size of the zenith-angle intervals (studied by dividing 
  the full zenith-angle range in four $\theta$ intervals with the same aperture), 
  the number of CIC cuts applied (using seven and three cuts instead of five), 
  the width of the CIC interval (employing a narrower muon range for the fit: 
  $\log_{10}N_\mu \approx [5.4, 6.0]$) and the size of the $N_\mu$-bins. 
  The total experimental error arising from the uncertainties in the CIC method 
  is found roughly between $-6\, \%$ and $+9\, \%$, while the corresponding
  MC error lies   between $\approx -7\, \%$ and 
  $\approx +8\, \%$. As we can see, both contributions are almost of 
  the same order of magnitude and constitute also an important source of 
  uncertainty in the estimation of the measured and predicted muon attenuation 
  lengths, respectively.
  
   \textit{Errors of the parameters of the muon correction function}. To
   evaluate the influence of this contribution on the final $\Lambda_\mu$ 
   results, we propagated the errors in the determination of the parameters 
   of the correction function (obtained under a mixed composition assumption
   with the QGSJET-II-02 model) to the $N_\mu$ data and hence to the muon 
   attenuation lengths. From table~\ref{tabAppB}, we observe that the 
   resulting shifts in the predicted and measured $\Lambda_\mu$ values are 
   in both cases small. Therefore this systematic source is not dominant.   

    \textit{Uncertainties in the spectral index of the primary cosmic ray spectrum}.
   Only the uncertainties of the MC based predictions take into account this
   source of systematic error, which is evaluated by using two different 
   values for the spectral index: $\gamma = -2.8$ and $-3.2$, in the simulated 
   data. The range of variation found in the corresponding $\Lambda_\mu$ results 
   with respect to the standard value with $\gamma = -3.0$ is quoted as the 
   systematic error from this contribution. In general, it results that the
   uncertainty in the spectral index has no major influence on the magnitude of 
   $\Lambda_\mu$ expected from the high-energy hadronic interaction models.
   
   \textit{Uncertainties in the primary composition}. Systematic uncertainties 
   for MC predictions include also the spreading of values when pure primary
   cosmic ray composition scenarios are considered. For these estimations, we 
   employed five distinct elemental primary nuclei: H, He, C, Si and Fe. On the 
   other hand, in order to reduce the influence of possible statistical effects, 
   we have increased, in each case, the size of the zenith-angle bins employed in 
   the CIC method. For this purpose, we reduced the number of $\theta$ intervals
   in the analysis. In particular, we employed four zenith-angle ranges, i.e. $\theta = 
   [0^\circ, 18.75^\circ],[18.75^\circ, 27.03^\circ], [27.03^\circ, 33.82^\circ]$
   and $[33.82^\circ, 40^\circ]$, all of them with approximately equal aperture. 
   We then extracted $\Lambda_\mu$ using the standard procedure for each primary 
   composition assumption. The biggest and smallest values of $\Lambda_\mu$ derived 
   in this way for each model were considered as the errors of the expected
   $\Lambda_\mu$ associated with the cosmic ray composition uncertainty. As we 
   can see from table~\ref{tabAppB}, they constitute the major source of 
   uncertainty in MC predictions. It is worth to point out that, for measured 
   data, this source of systematic error is already taken into account. 
   Specifically, it is considered when calculating the contribution to the total 
   experimental uncertainty due to the systematic biases of the corrected $N_\mu$ 
   for each of the aforementioned primary nuclei.

   \section{Further systematic checks}
   \label{appB}

   \renewcommand{\thefigure}{C\arabic{figure}}
   \setcounter{figure}{0}

    \setcounter{table}{0}
    \renewcommand{\thetable}{C\arabic{table}}

	 In this part of the paper, we evaluate the influence of suspected 
   sources of systematic errors that might be at work in this analysis.\\

   \textit{Aging of the muon detectors}
    From the experimental point of view, one of these possibilities is the 
   natural aging of both the plastic scintillator detectors and the PMT's
   of the KASCADE muon detectors. To quantify this effect,
   the measured data was divided in three subsamples with effective
   observation times of approximately the same order of magnitude and
   ordered in time. For each subset of data, the muon attenuation length
   was estimated (table~\ref{tabAppC}).
   No dependency of the measured $\Lambda_\mu$ on the time is
   observed. All values for the three different periods are in very good 
   agreement within their own errors and are in accordance 
   with the mean value shown in table~\ref{tab1} for the whole measured data 
   sample (considering only statistical uncertainties, deviations are between 
   $\approx - 0.25 \, \sigma$ and $\approx + 0.29 \, \sigma$). In consequence, 
   it can be concluded that the aging of the muon detectors is not  
   responsible for the observed discrepancy between the measured and the 
   predicted muon attenuation lengths.

   \subsection{Evolution of the elemental abundances of cosmic rays}
    
    As we know from  detailed studies performed in \cite{KGPRD, KGPRL, KGunfold}, 
   the chemical composition of cosmic rays in the energy interval analysed is
   changing from light to heavy. Therefore the actual event samples contain
   a wide range of early and late developing showers, which might lead to 
   a significant increase of $\Lambda_\mu$ in comparison to the results
   with a single or equal-abundance composition scenarios. To quantify
   the influence of this effect, we used a toy model for the elemental 
   composition of cosmic rays between $10^{16}$ and $10^{18} \, \mbox{eV}$
   following the results of \cite{KGPRD, KGPRL, KGunfold}. The model included 
   the spectral features observed in the light and heavy components. Using the
   data from QGSJET-II-02 along with this elemental abundances, we calculated 
   $\Lambda_\mu$. The result was just  $1.4 \, \%$ smaller than the one
   obtained for the mixed composition assumption based on equal abundances.
   Therefore, the changing elemental abundances of cosmic rays in the
   studied energy regime is not causing the observed anomaly. 

 \begin{table}[!t]
    \begin{center}
    \caption{$\Lambda_\mu$ measured for different KASCADE-Grande subsets
    of data corresponding to three distinct periods. Statistical and
    systematic errors are shown in order of appearance. The latter only
    contains the contribution from the global fit.}
   \begin{tabular}{l|c|c|c}
   \hline 
    &Period & Effective time (s) & $\Lambda_\mu$($\mbox{g}/\mbox{cm}^{2}$)\\ 
    \hline
    Sample 1 & $20/12/2003 - 07/11/2006$ & $3.3 \times 10^{7}$ & $1233 \pm 115 \pm 89$\\
    Sample 2 & $07/11/2006 - 11/04/2009$ & $5.2 \times 10^{7}$ & $1295 \pm 105 \pm 85$\\
    Sample 3 & $11/04/2009 - 31/10/2011$ & $3.9 \times 10^{7}$ & $1219 \pm 120 \pm 89$\\
   \hline
   \end{tabular}
   \label{tabAppC}
   \end{center}
  \end{table}

   \subsection{Fluctuations on the number of registered muons per station}
\begin{figure}[!t]
 \centering
 \includegraphics[width=3.2in]{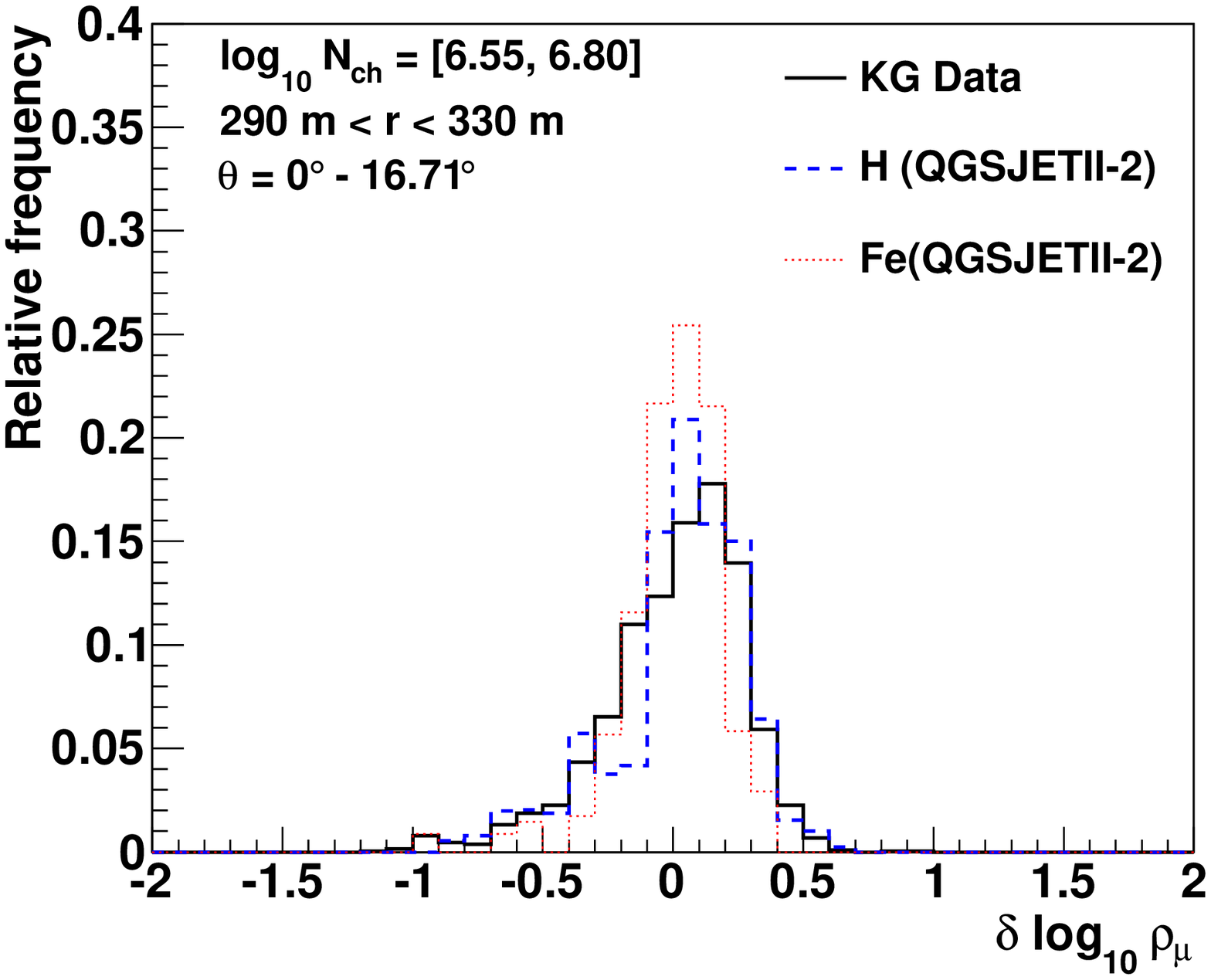}
 \caption{Measured distribution of the $\rho_\mu$ fluctuations for  
 a radial interval $[290 \, \mbox{m}, 330 \, \mbox{m}]$ at the shower plane
 and for vertical showers in the range $\log_{10} N_{ch} = [6.55, 6.80]$,
 where $N_{ch}$ has been corrected for attenuation effects in the atmosphere
 and normalized at $\theta = 22^{ \circ}$ with the CIC method. The
 distribution is compared with the predictions of QGSJET-II-02 for 
 primary protons and iron nuclei.}
 \label{RhoMuFluctuations}
\end{figure}

   Another interesting possibility is the influence of fluctuations 
  on the number of registered muons  $n_\mu$ per KASCADE detector. The number of 
  muons collected by a muon station is in general small, therefore fluctuations 
  may play an important role here. In addition, fluctuations from MC simulations
  for $n_\mu$ might be different from the experimental ones. All these effects
  together may lead to a bias in the reconstructed $N_\mu$ explaining the 
  observed $\Lambda_\mu$ deviations. In order to find out whether fluctuations 
  on $n_\mu$ are responsible for the deviations, QGSJET-II-02 simulations were 
  employed. First, $\rho_\mu$ fluctuations were obtained from the distributions
  of the density of muons as a function of the distance to the 
  core at the shower plane (see as an example, fig.~\ref{RhoMuFluctuations}). 
  The muon densities, $\rho_\mu(r)$, were built event-by-event by dividing the 
  EAS plane in concentric rings ($20 \, \mbox{m}$  width each) and then by dividing, 
  for each radial interval, the corresponding amount of detected muons by the sum 
  of projected effective areas of the active detectors located in that particular bin. 
  
   Fluctuations were extracted from both, MC and experimental data for the 
  different zenith-angle ranges and for several $N_{ch}$ intervals, 
  where $N_{ch}$ was corrected for attenuation effects in the atmosphere using the 
  CIC method. To separate the data, the charged number of particles was chosen  
  instead of $N_\mu$  because in the former both the observed resolution and the 
  agreement between the corresponding measured attenuation length and the MC 
  predictions are better. MC fluctuations were obtained only for proton and iron 
  nuclei as primaries, respectively. For experimental data, fluctuations might be 
  overestimated since they might contain contributions from different primary elements.
   Once fluctuations were calculated, they were applied with a simulation program
  event-by-event to the MC data sets to estimate the number of particles detected
  per KASCADE muon station per simulated shower under each of the above fluctuation
  scenarios. For a given MC event 
  with true muon content $N_\mu$, the number of muons hitting each KASCADE muon 
  station is estimated according to the geometry of the station and the 
  muon lateral distribution function of equation (\ref{eq1}). For this
  estimation the true values of the shower core position and arrival direction
  are needed. They are taken from the input parameters used in CORSIKA to 
  simulate the shower. Once the number of muons per station is known, this 
  quantity is allowed to fluctuate using the corresponding statistical 
  distributions obtained from the experiment or simulated data. Then, the 
  new set of $n_\mu$ values are stored and the mean deposited energy per 
  muon station is estimated. Henceforth, the standard KASCADE-Grande 
  reconstruction software is applied.  The muon attenuation lengths are
  finally obtained from the  reconstructed MC data sets using the standard 
  procedure described in section \ref{lambdamuon}. 

  Interestingly, the final results with MC simulations showed that
  the $\Lambda_\mu$ value obtained with experimental fluctuations stays 
  above the corresponding result derived when using the MC ones, however 
  the differences are small, just below $15\, \%$ for QGSJET-II-02. In 
  consequence, the effect of the fluctuations on the number of muons 
  per KASCADE muon station can not explain the observed $\Lambda_\mu$ 
  discrepancy between measured and predicted data.

 \subsection{Uncertainties of air shower parameters}
   The influence of systematic errors coming from uncertainties in the 
  reconstruction of the core position, arrival direction and the number of 
  muons per detector from the deposited energy were also studied. 
  For this purpose, new MC data sets were generated based on QGSJET-II-02 
  and by using the true shower location, arrival 
  direction and number of muons hitting the KASCADE muon detectors
  in the reconstruction stage of the MC events. This way, the $N_\mu$ 
  estimated for the resulting data sets has no influence from the 
  systematic errors due to mislocation of the core, misalignment of the 
  reconstructed shower axis or wrong estimation of $n_\mu$ per station. 
  For the new data sets, the $\Lambda_\mu$ values are extracted and
  are compared with the corresponding attenuation lengths from the data
  where the uncertainties on the shower parameters are considered
  (for simplicity, in both cases, no muon correction function was applied).
  From the comparison, it is concluded that the effect of the abovementioned
  systematic errors on $\Lambda_\mu$ is to modify its magnitude, but by 
  a negligible amount ($\lesssim 3 \, \%$). 
  
  These are conservative predictions associated with
  the effects of core and angular resolutions. One can ask what would
  happen if the actual magnitude of systematic errors of the core 
  position and arrival direction were somewhat different. For this
  possibility there is not much room left, since the resolution of 
  the Grande detector has been checked out with the KASCADE array, 
  which works independently of the former as mentioned in section 
  \ref{Setup}. By introducing these errors \cite{KG} in our MC 
  simulations the muon attenuation length varies just within
  $7 \, \%$. Therefore, in light of the above results, it is
  unlikely that systematic errors due to shower core position and 
  arrival direction could be the main cause for the $\Lambda_\mu$ 
  deviation between experimental data and MC expectations.

   \subsection{Uncertainties of the muon LDF}
   \label{appC5}
   The fact that there is an intriguing dependence of the muon  
  attenuation length on the core position, which is not predicted by 
  simulations, suggests the presence of another source of systematic 
  error of $\Lambda_\mu$ (see \ref{appA}). One possibility could be 
  found at the shape of the muon lateral distribution function. During 
  reconstruction the slope of the LDF is kept constant due to the fact 
  that the KASCADE muon detectors only sample a limited portion of the 
  EAS. However, it is known that, although the measured LDF for muons is 
  bracketed by simulation results, the observed slopes are different 
  from MC predictions \cite{KGMudensities}.  By comparing the 
  slopes of the mean muon lateral distribution functions expected from
  MC simulations and the ones observed in experimental data, we found 
  that, on average, the MC distributions are flatter than the measured 
  ones. These differences clearly suggest the presence of a potential 
  source of systematic error of $N_\mu$, which may be also contributing 
  to the observed anomaly. To estimate the possible contribution from 
  this effect to the systematic error of $\Lambda_\mu$, first, for each 
  zenith-angle interval and the $\log N_{ch}^{CIC}$ range discussed 
  in section \ref{density}, we fitted the QGSJET-II-02 and the 
  experimental mean muon density distributions with formula (\ref{eq1}) but
  using $p_1$ and $N_\mu$ as free parameters.  This was done in order 
  to get an estimation of the flatness of the muon density distributions 
  and to quantify the differences between the slopes of the 
  experimental and the expected LDF's. The fits were performed on the 
  radial interval $r > 160 \, \mbox{m}$. For MC, we applied the fits on the 
  data sets for pure elements and mixed composition. From the fitted 
  values of  $p_1$, it was found that, in general, the actual mean muon 
  radial density distributions are on average $7 \, \% \pm 15 \, \%$ 
  steeper than the MC simulated ones. 
  
  To evaluate the effect of using a flatter muon LDF to fit our data,
  we considered the MC data sets of QGSJET-II-02 for a mixed composition 
  scenario and proceeded to reconstruct $N_\mu$ event-by-event with a flatter 
  muon LDF. The latter was performed by decreasing the magnitude of
  $p_1$ by $22\, \%$ in the LDF formula employed for the standard EAS
  reconstruction, see eq. (\ref{eq1}). This percentual decrement corresponds to
  the upper limit of the $1 \, \sigma$ interval found for the difference
  between the $p_1$ values of the MC and measured muon LDF's. For the
  above variation, we found that  $\Lambda_\mu$ is shifted by $+6 \, \%$. 
  In addition, the dependence of $\Lambda_\mu$ on the distance to the
  KASCADE muon cluster became larger than the one observed in table~\ref{tabAppB}
  for QGSJET-II-02. In particular, $\Lambda_\mu$ decreases
  by  $-7 \, \%$ for events with cores between $R = 270 \, \mbox{m}$ to 
  $360 \, \mbox{m}$, and increases by $+6 \, \%$ at farther distances
  ($R = [360, 440] \, \mbox{m}$). If now the magnitude of the parameter 
  $p_1$ of formula (\ref{eq1}) is increased by $22\, \%$, in order to have 
  a steeper muon LDF as suggested by the measured data, then we observe 
  that the experimental  $\Lambda_\mu$ is reduced only by $\approx 8 \, \%$ 
  ($\sim 99 \, \mbox{g}/\mbox{cm}^{2}$), while the core dependence 
  of $\Lambda_\mu$ remains still high ($\pm 9 \, \%$). Therefore, we see that 
  the systematic errors of $\Lambda_\mu$ are not enough to be the cause of the
  discrepancy. 
  
  To give a better estimation of the effect of the $\Delta p_1$
  differences between the measured and the MC data and with the aim of 
  confirming the conclusion of the previous analysis, we used an alternative 
  approach: we weighted the $\rho_\mu(r)$ distributions of the QGSJET-II-02 
  events for the mixed composition assumption to reproduce a steeper LDF 
  in closer agreement with the one observed from the measurements. Then we 
  applied the standard KASCADE-Grande reconstruction algorithm to the 
  aforementioned MC events to obtain $N_\mu$ from which we calculated 
  $\Lambda_\mu$. Finally, the latter is compared with the standard result 
  obtained from the unmodified data sets. The weight was applied by 
  multiplying the number of events recorded in each station by the factor 
  $(r/320 \, \mbox{m})^{\Delta p_1}$, where $\Delta p_1 = p_1^{KG} - p_1^{MC}$ 
  is the mean difference in $p_1$ obtained from the study described in the previous 
  paragraph. Since $\Delta p_1 = -0.07 \pm 0.16$,
  we used the lower limit of this interval for the estimation of the $\Lambda_\mu$ 
  systematic uncertainty. The result was an increase of $\approx +8 \, \%$ 
  ($\sim 57 \, \mbox{g}/\mbox{cm}^{2}$), which is of the order of magnitude of 
  the systematic error already calculated in the aforementioned paragraph.
  
   One may argue that the individual differences between the LDF's at 
   different zenith-angles may be contributing in some way to the 
   $\Lambda_\mu$ systematics too. In general, we have observed that
   both the MC and measured mean muon radial density distributions become
   flatter as the zenith-angle increases. However, the slope of the
   measured LDF's decreases faster than that derived from MC simulations. 
   To quantify the influence of these effects on the muon anomaly, first
   we modelled the above differences based on the observed $\Delta p_1(\theta)$ as 
   obtained for the interval $\log N_{ch}^{CIC} = [7.04, 7.28]$. The differences 
   were derived by comparing the experimental data with the results from the 
   QGSJET-II-02 model for a mixed composition scenario and primary spectrum
   $\propto E^{-3}$. Then we weighted the muon LDF's from the QGSJET-II-02 data 
   sets by using the factor $(r/320  \, \mbox{m})^{\Delta p_1(\theta)}$, with 
   $\Delta p_1(\theta) = -0.138 + 0.143 \cdot \theta$, with $\theta$ in radians. 
   Finally, we reconstructed $N_\mu$ event-by-event and obtained 
   $\Lambda_\mu$ by the usual procedure. The result was a shift of 
   $\sim + 2 \, \%$ on the simulated $\Lambda_\mu$.  
	
	In summary,  we conclude that it is improbable 
   that the uncertainty on the slope of the LDF is the main cause of the 
   deviation on the muon attenuation length.
   
   \subsection{Influence of the muon correction function}
    \label{appC6}
    The prime suspect behind the $\Lambda_\mu$ anomaly is the muon correction 
  function applied to the data. In general, the effect of this function on 
  the estimated $\Lambda_\mu$ is to shift its magnitude by $+13 \, \%/-3 \, \%$ 
  for MC simulations and $+ 17 \, \%$ for experimental data with 
  respect to the value extracted from the uncorrected $N_\mu$.
  It is observed that the amount of shift for the experimental 
  value is bigger than that for MC estimations. However, it
  does not explain the discrepancy. In fact, a more detailed
  analysis based on the mean lateral muon densities (see section
  \ref{density}) revealed that the differences between the 
  measured and expected muon attenuation lengths are not an artefact 
  from the application of the muon correction function on the data. 
  In particular, it was observed that they can be tracked down to 
  differences between the experimental and predicted evolutions of
  the local mean muon densities in the shower front with the angle $\theta$.
  This asseveration can be probed by modifying in an
  artificial way the zenith-angle evolution of the muon lateral 
  distribution functions obtained from MC simulations. We have 
  employed the same simulated MC data sets used to study the 
  impact of the uncertainties in the slope of the muon LDF's 
  on $\Lambda_\mu$, and we have multiplied the corresponding muon
  densities by the factor  $\left[ e^{X_0 (1 - \sec{\theta}) \cdot
  (1/\bar{\alpha}^{KG}_\mu  - 1/\bar{\alpha}^{MC}_\mu)}
   \cdot (r/320   \, \mbox{m})^{\Delta p_1(\theta)} \right]$. Here, 
  $\bar{\alpha}^{KG}_\mu = (1159 \pm 110)  \, \mbox{g}/\mbox{cm}^{2}$
  is the average value of the muon absorption length for the
  experimental data in the radial interval $r = [220 \, \mbox{m},
  380 \, \mbox{m}]$ and the shower size range $\log N_{ch}^{CIC} = 
  [7.04, 7.28]$ (see fig. \ref{Results_absorption_New}). On the
  other hand, $\bar{\alpha}^{MC}_\mu = (821 \pm 28)  \, 
  \mbox{g}/\mbox{cm}^{2}$ is the corresponding value for the
  QGSJET-II-02 based simulations (mixed composition data in
  fig. \ref{Results_absorption_New}). After applying the full
  reconstruction procedure to the new simulated data, we found that
  $\Lambda_\mu^{MC} = (1116 \pm 184) \, \mbox{g}/\mbox{cm}^{2}$, which
  is in pretty good agreement with the measured value. 
  
  \subsection{Fluctuations on the local values of atmospheric
  temperature and pressure}
  \label{appC7}

   The influence of local
  variations of the air pressure and temperature on our results were 
  investigated. At the site, the mean pressure at ground during the
  DAQ period used for our analysis was 
  $\bar{P} = 1003.0 \pm 8.5 \, \mbox{mbar}$, which is pretty close 
  (within the experimental RMS variations) to the nominal value of 
  $\approx 1002.2 \, \mbox{mbar}$ 
  ($P_0 = 1022 \, \mbox{g}/\mbox{cm}^2$) used for the MC simulations.   
  To evaluate the influence of this small difference in the measured
  $\Lambda_\mu$, data within a small interval $\Delta P_0$ around $P_0$
  was chosen and the corresponding  muon attenuation length was evaluated. 
  In particular, we used  $\Delta P_0 = [998.3 \, \mbox{mbar}, 
  1006 \, \mbox{mbar}]$. 
  This range was selected in such a way that $P_0$ coincides with the
  median of the pressure distribution for the corresponding interval.
  The result for $\Lambda_\mu$ is shown in table~ \ref{tabPT}.
  This value is just $0.008 \, \sigma$ (for statistical errors only) below 
  that corresponding to the full experimental data set. Therefore,  
  the difference between the values $P_0$ in the interval selected and 
  $\bar{P}$ can not be the main cause of the $\Lambda_\mu$ discrepancy.

   To go further, we investigated the effect of the tails of
  the $P$ distribution. For this purpose, we considered
  two additional data sets: one with $P > 1006 \, \mbox{mbar}$ and 
  another one with $P < 998.3 \, \mbox{mbar}$, and we calculated
  $\Lambda_\mu$ for each case. The extracted values are presented in 
  table \ref{tabPT}. They are within $-0.4 \, \sigma$ and $+0.9 \, 
  \sigma$ (using only statistical uncertainties for the comparison), 
  respectively, from the main result obtained for the 
  whole KASCADE-Grande data set. The magnitude of these deviations 
  can not explain the observed anomaly of the muon attenuation
  length. If the smallest value of $\Lambda_\mu$ obtained
  from the present analysis with different $P$ intervals 
  is compared with the MC predictions of table \ref{tab1}, 
  deviations from $+3.2\, \sigma$ to $+4.6 \, \sigma$ arise 
  (employing only statistical errors).
  
   On the other hand, it is also worth mentioning that a possible hint
  for a dependence of the $\Lambda_\mu$ discrepancy  with the mean
  atmospheric pressure seems to be observed  in the data
  (see table \ref{tabPT}). In particular, the results seem to suggest
  that the disagreement between the measured and predicted $\Lambda_\mu$
  parameters grows when decreasing the mean value of $P$. The effect 
  seems to be the result of
  an apparent reduction in the estimated number of muons at lower
  pressures, which is more important for vertical showers.
  For example, when comparing
  the muon attenuation curves derived for the data sets with $P < 998.3  \, 
  \mbox{mbar}$ and $P > 1006  \, \mbox{mbar}$, respectively, at the same CIC cut:
  $\log_{10} [J/ (\mbox{m}^{-2} \cdot \mbox{s}^{-1} \cdot \mbox{sr}^{-1})] = -8.60$, 
  it is observed that for showers closer to the zenith (first angular bin), the 
  magnitude of $N_\mu$ derived from the CIC method for the interval with highest 
  $P$ is $\approx 4.5 \, \%$ bigger than that obtained for the interval of 
  lowest  atmospheric pressure, while for inclined showers (last zenith-angle bin) 
  the difference is negligible and it amounts to $\approx 0.7 \, \%$. The
  interpretation of the results given here is still tentative as the statistical 
  errors for the subsamples of table \ref{tabPT} are not small.

      \begin{table}[!t]
    \begin{center}
    \caption{Attenuation lengths for the muon number extracted 
    from experimental data for different intervals of pressure, $P$ (mbar), 
    and temperature, $T$ (${}^{\circ} C$), at the site. Statistical and
    systematic errors are shown in order of appearance. The latter only
    contains the contribution from the global fit.}
   \begin{tabular}{l|c|c|c}
   \hline 
    Interval & Mean $(P,T)$ & Effective time (s) & $\Lambda_\mu$($\mbox{g}/\mbox{cm}^{2}$)\\ 
    \hline
    $P > 1006.0$          & ($1012.0 \pm 4.4$, $7.8 \pm 7.7$) & $5.24 \times 10^{7}$ & $1204 \pm 104 \pm 79$\\
    $P = [998.3, 1006.0]$ & ($1002.0 \pm 2.1$, $12.9 \pm 7.5$)& $5.24 \times 10^{7}$ & $1255 \pm \, 99\pm 81$\\
    $P < 998.3$           & ($ 992.5 \pm 5.5$, $9.8 \pm 7.4$) & $3.43 \times 10^{7}$ & $1405 \pm 139\pm 109$\\
    &&&\\ 
    $T > 14.15$           & ($1002.0 \pm 5.4$, $19.4 \pm 4.0$) & $4.54 \times 10^{7}$ & $1249 \pm 111\pm 84$\\
    $T = [6.45, 14.15]$   & ($1003.0 \pm 8.2$, $10.3 \pm 2.2$) & $4.69 \times 10^{7}$ & $1234 \pm 124\pm 86$\\
    $T < 6.45$            & ($1005.0 \pm 10.6$, $1.5 \pm 3.5$) & $4.68 \times 10^{7}$ & $1310 \pm 160\pm 88$\\
   \hline
   \end{tabular}
   \label{tabPT}
   \end{center}
  \end{table}

   Regarding the influence of the local variations of temperature
  on $\Lambda_\mu$, we have found that it is not significant. The temperature at the
  site was continuously monitored from the top of a tower at $200 \, \mbox{m}$ 
  above the ground. From the records of the temperature during the DAQ 
  period of the analysed data, we found that the mean value of the 
  temperature was $\bar{T} = {10.27}^{\circ} \, C$ with a standard deviation 
  of ${7.88}^{\circ} \, C$. To study the effect of the local temperature 
  variations on the muon attenuation length, we divided our data in three 
  subsets according to the following temperature intervals: $T < {6.45}^{\circ} \, C$,
  $T = [{6.45}^{\circ} \, C, {14.15}^{\circ} \, C]$ and $T > {14.15}^{\circ} \, C$, 
  each of them with approximately the same statistics. Then we applied our
  standard analysis to find $\Lambda_\mu$ in each case (table \ref{tabPT}). 
	The results show variations from $- 0.1 \, \sigma$ to $+ 0.3 \, \sigma$
  from the measured value reported in table \ref{tab1} for the whole
  experimental data set  (comparisons were performed using only statistical uncertainties). 
  Therefore, it is unlikely that the variations in the local temperature could be 
  the cause of the observed $\Lambda_\mu$ anomaly.

 \section{The attenuation length for $N_{ch}$}
 \label{Lambdach}

 \renewcommand{\thefigure}{D\arabic{figure}}
  \setcounter{figure}{0}

\setcounter{table}{0}
\renewcommand{\thetable}{D\arabic{table}}

\begin{figure}[!t]
 \centering
 \includegraphics[width=3.2in]{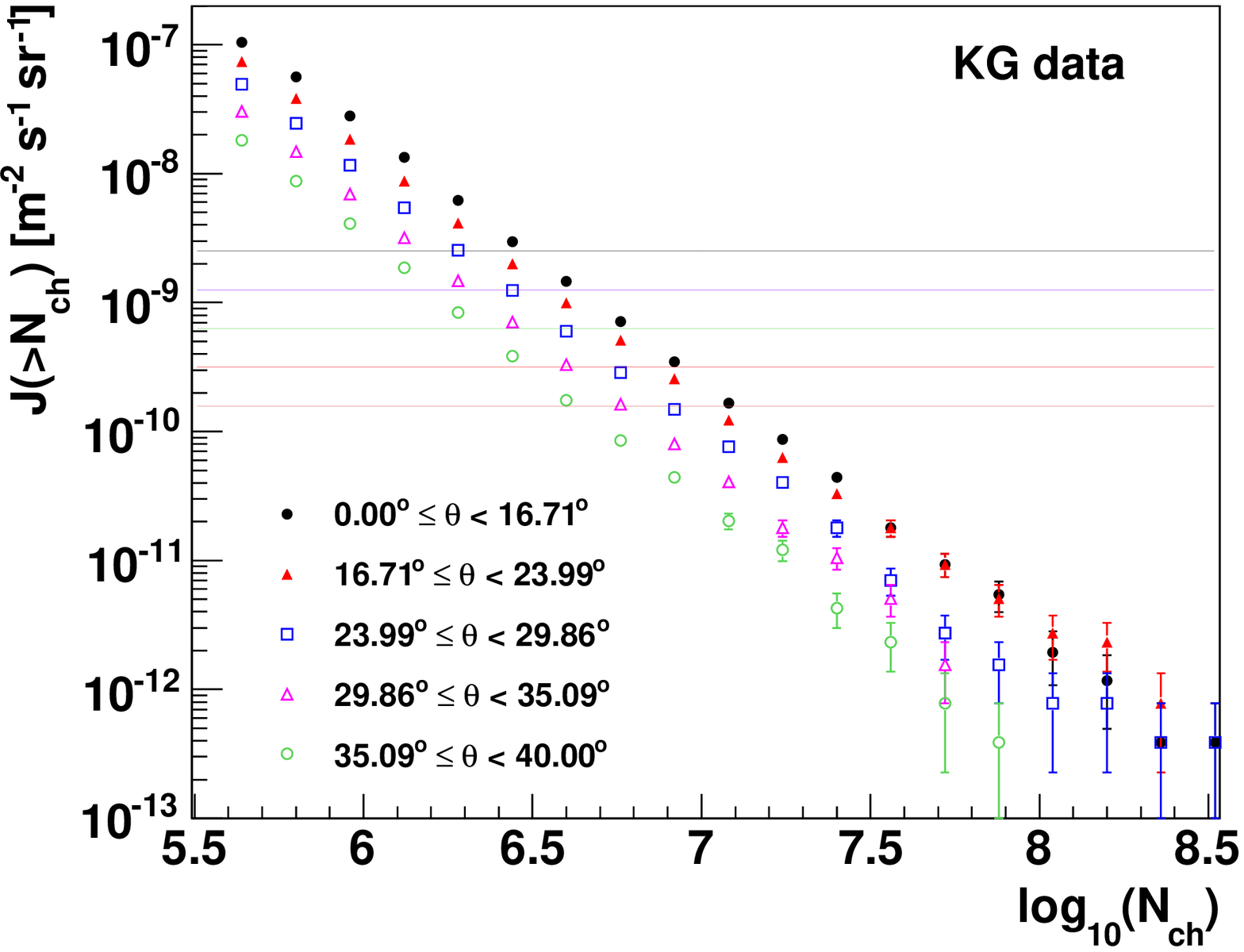}
 \includegraphics[width=3.2in]{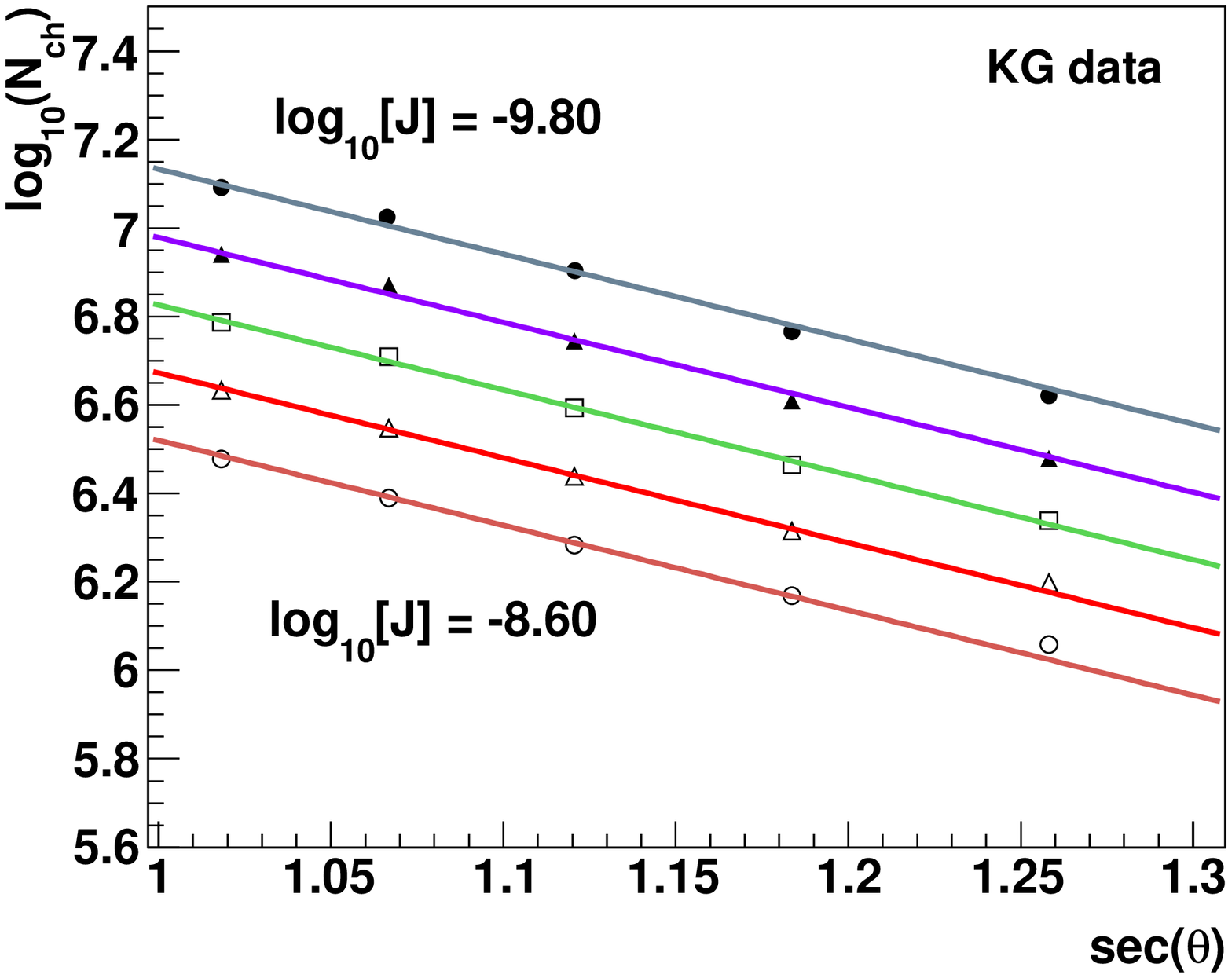}
  \caption{Left: $N_{ch}$ integral spectra for five zenith-angle intervals 
  derived from the measurements with the KASCADE-Grande observatory. 
  Error bars represent statistical uncertainties. The CIC cuts employed 
  in this work are shown as horizontal lines. Right: $N_{ch}$ attenuation 
  curves obtained by applying several constant intensity cuts to the 
  KASCADE-Grande integral spectra, $J_{ch}$. The cuts decrease from the 
  bottom to the top in units of $\Delta \log_{10} [J/(\mbox{m}^{-2} \cdot 
  \mbox{s}^{-1} \cdot \mbox{sr}^{-1})] = -0.30$. Errors are smaller than 
  the size of the symbols. They take into account statistical uncertainties, 
  errors from interpolation as well as the correlation between adjacent 
  points when interpolation was applied.}
  \label{JchandAttCurves}
\end{figure}

 \begin{figure}[!b] 
 \centering
 \includegraphics[width=3.2in]{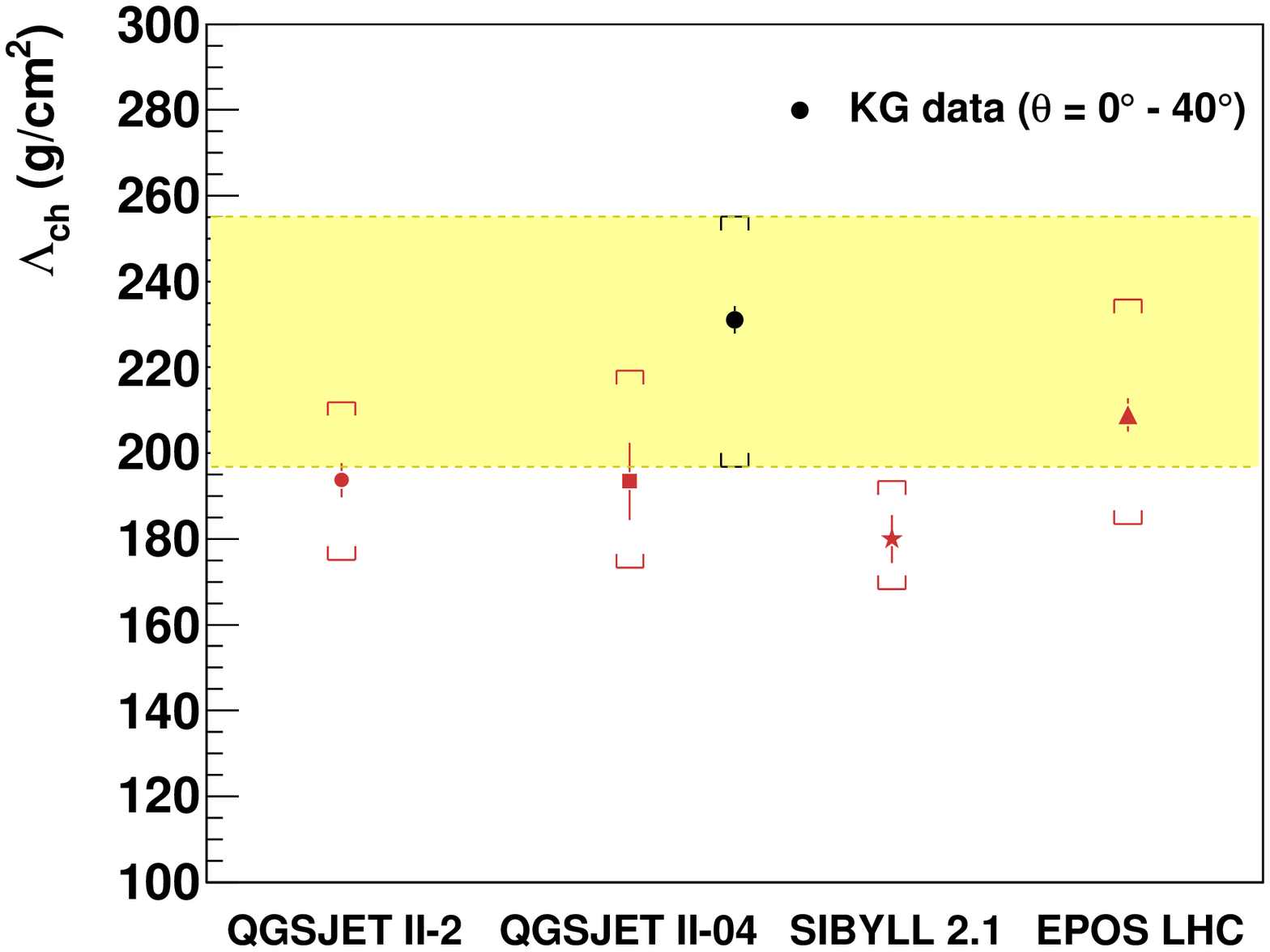}
  \caption{$N_{ch}$ attenuation lengths extracted from Monte Carlo 
  (lower points) and experimental data (upper black 
  circle). Error bars indicate statistical uncertainties, while
  the brackets represent the total errors (systematic plus statistical 
  errors added in quadrature). The shadowed band covers the total 
  uncertainty estimated for the experimental result.}
 \label{LambdaNchKGvsMC}
\end{figure}

  In order to complement the present study,  a last check was
 performed, but on $N_{ch}$, which includes the number of muons 
 and electrons of the shower. In this check, the $N_{ch}$ 
 attenuation length, $\Lambda_{ch}$, was estimated from the
 KASCADE-Grande measurements of air showers and the result
 was compared with the predictions from the hadronic 
 interaction models of section \ref{MC}. The extraction procedure 
 of $\Lambda_{ch}$ was identical to the one employed with $\Lambda_\mu$, 
 with the only exception that no correction function was applied. 
 The latter was not necessary for the analysis, since in KASCADE-Grande
 the charged particle content of EAS is determined with a better 
 precision than the muon number\footnote{In turn, the number of 
 electrons can be estimated even with a better precision than 
 $N_{ch}$ in KASCADE-Grande. For example, for our data set, after 
 applying quality cuts, MC predictions indicate that for shower 
 sizes $\leq 3.2 \times 10^{8}$, the systematics on $N_e$ are 
 $\lesssim 7\, \%$, while for $N_{ch}$ are $\lesssim 12\, \%$.} \cite{KG}.

     \begin{table}[!t]
    \begin{center}
    \caption{Attenuation lengths for the charged particle number extracted 
    from Monte Carlo and experimental data. $\Lambda_{ch}$ is presented along 
    with their statistical and systematic errors (in order of appearance). Also 
    the deviations (in units of $\sigma$) 
		of the measured $\Lambda_{ch}$ 
    from the predictions of different hadronic interaction models are shown. 
		The
    one-tailed confidence levels (CL) that the measured value is in 
    agreement with the MC predictions are also presented.}
   \begin{tabular}{l|c|c|c|c|c}
   \hline 
    &  QGSJET-II-02 & QGSJET-II-04 & SIBYLL 2.1  & EPOS-LHC & KG data \\ 
   \hline
   &&&&&\\
   {\bf $\Lambda_{ch}$ $(\mbox{g/cm}^2)$} 
   &$194\pm 4^{+ 18}_{-18}$ 
   &$193\pm 9^{+24}_{-18}$ 
   &$180\pm 6^{+ 12}_{-10}$ 
   & $209\pm 4^{+27}_{-\, 25}$
   & $231\pm 3^{+24}_{-34}$\\ 
   &&&&&\\
   \hline
   Deviation ($\sigma$) & $+0.96$ & $+0.88$ & $+1.39$ & $+0.51$ &\\ 
   CL (\%)              & $16.78$ & $19.02$ & $ 8.25$ & $30.56$ &\\
   \hline
\end{tabular}
   \label{tab3}
   \end{center}
  \end{table}

  The measured $N_{ch}$ integral spectra upon which the analysis is 
 performed are presented on the left side of fig.~\ref{JchandAttCurves}  
 along with the applied CIC cuts. On the right side of the same 
 figure, the $N_{ch}$ attenuation curves extracted with the CIC method 
 are also shown. As before, the $\Lambda_{ch}$  is obtained from a 
 global fit with a relationship like (\ref{eq3}) to the measured attenuation 
 curves. The resulting value is plotted on fig.~\ref{LambdaNchKGvsMC} 
 together with the predictions from QGSJET-II-02, SIBYLL 2.1,
 QGSJET-II-04 and EPOS-LHC. The magnitudes of the measured 
 and predicted values of $\Lambda_{ch}$ are displayed in table 
 \ref{tab3}.

  \begin{table}[!t]
    \begin{center}
    \caption{Total uncertainties on the predicted and experimental
    $\Lambda_{ch}$. The different contributions from the
    systematic and statistical errors are also shown.}
   \scriptsize
   \begin{tabular}{l|c|c|c|c|c}
   \hline 
    &  QGSJET-II-02 & QGSJET-II-04 & SIBYLL 2.1 & EPOS-LHC & KG data \\ 
   \hline
   \textbf{Statistical error ($\%$)}  &&&&&\\    
   Statistical fluctuations      &$\pm 2.05$ &$\pm 4.65$ &$\pm 3.10$ &$\pm 1.89$ &$\pm 1.38$\\
   &&&&&\\    
   \textbf{Systematics ($\%$)}  &&&&&\\
   Nch systematics               &$-1.90$    &$-2.58$    &$+0.94$    &$-9.75$    &$-13.55/+8.79$\\
   Global fit                    &$\pm4.34$  &$\pm4.67$  &$\pm4.58$  &$\pm4.71$  &$\pm4.94$\\
   Core far from KASCADE         &$+2.26$    &$-0.60$    &$+0.54$    &$+0.79$    &$+2.04$\\
   ($R = [360, 440] \, \mbox{m}$)&&&&&\\
   Core close to KASCADE         &$-2.91$    &$-0.12$    &$-1.99$    &$-1.10$    &$-2.07$\\
   ($R = [270, 360] \, \mbox{m}$)&&&&&\\
   Bin size                      &$-1.90$    &$+1.02$    &$+2.81$    &$-2.46$    &$-1.59$\\ 
   Narrower CIC interval         &$-0.75$    &$-0.42$    &$-0.73$    &$+0.41$    &$-1.49$\\
   ($\log_{10}N_{ch} \approx [6.1, 7.2]$) &&&&&\\
   Broader zenith-angle interval &$+0.01$    &$-0.33$    &$+1.25$    &$+0.28$    &$-0.71$\\
   (Four angular bins)&&&&&\\
   Number CIC cuts               &$-0.32/+0.23$ &$-0.10/+0.09$ &$-0.96/+0.27$    &$-1.30/+0.20$    &$-0.08/+0.47$\\
   Spectral index uncertainties  &$-1.11/+3.28$ &$-0.44/+2.12$ &$-0.56/+1.13$    &$-1.89/+1.46$    &$-$\\
   ($\Delta \gamma = \pm 0.2$) &&&&&\\
   Composition                   &$-7.18/+6.99$ &$-7.61/+11.37$&$-2.36/+3.72$    &$-3.98/+11.68$    &$-$\\
    \textbf{Total (\%)}   &&&&&\\
                                 &$+9.37$   &$+13.35$    &$+7.52$   &$+12.86$    &$+10.39$\\
                                 &$-9.61$   &$-10.43$    &$-6.47$   &$-12.22$    &$-14.81$\\
    
   \hline
   \end{tabular}
   \label{tabAppD}
   \end{center}
  \end{table}

   To investigate the agreement between the measurement 
 and the predictions from MC simulations, a simple statistical analysis 
 was applied. Deviations of the experimental $\Lambda_{ch}$ from 
 estimations of the models were computed and confidence levels 
 for agreement with the predictions of the hadronic interaction 
 models were derived. The results are presented in table~\ref{tab3}. 
 Herein a good consistency between experimental data and the 
 predictions of the high-energy hadronic interaction models
 can be seen, since the statistical analysis gives deviations between 
 $+0.51 \, \sigma$ and $+1.39 \, \sigma$, with a CL from $8.25 \, \%$ to 
 $30.56 \, \%$, respectively, which are as a matter of fact quite 
 satisfactory.

 The total uncertainties of $\Lambda_{ch}$ are presented in table
 \ref{tabAppD} along with their corresponding statistical and
 systematic errors. All of them were calculated in the same way 
 that for $\Lambda_{\mu}$. The results were found to vary in the range
 from $\approx -15 \, \%$ to $\approx +13\, \%$. In experimental data, 
 an important contribution to the total error of $\Lambda_{ch}$ 
 (between $\approx -13 \, \%$ and $\approx +9\, \%$) is the systematic 
 uncertainty of $N_{ch}$. The latter was estimated from MC simulations 
 and confirmed with experimental investigations \cite{KG}. On the other 
 hand, in contrast to the $\Lambda_\mu$ case, here no relevant dependence 
 of the measured $\Lambda_{ch}$ with the radial distance was found,
 for the corresponding variations of $\Lambda_{ch}$ were
 within $\pm 2 \, \%$ (see table ~\ref{tabAppD}). The reason is that,
 for the charged component of EAS the LDF is well measured event-by-event 
 across the Grande detector area. Regarding MC
 simulations, a sizeable contribution in this case came from the
 uncertain knowledge of the primary composition of the
 experimental sample. This was estimated from the data sets for
 the pure and mixed composition scenarios (as in
 the case of $\Lambda_\mu$). It resulted that this source of
 uncertainty has a contribution from $-8 \, \%$ to $+12 \, \%$
 to the total MC error depending of the hadronic interaction model.


\begin{thebibliography}{00}
\bibitem{MCOstap} S. Ostapchenko,  Czech. J. Phys. 56 (2006) A149.
\bibitem{MCPierog} T. Pierog, R. Engel, D. Heck, Czech. J. Phys. 56 (2006) A161.
\bibitem{KASCADE-model1} W.D. Apel et al., KASCADE-Grande Collaboration, Astrop.
Phys. 24 (2005) 1.
\bibitem{Cazon} L. Cazon, R.A. Vazquez, A.A. Watson, E. Zas, Astrop. Phys. 21 (2004)
71; L. Cazon, R.A. Vazquez, E. Zas, Astrop. Phys. 23 (2005) 393.
\bibitem{Meurer} C. Meurer, J. Bluemer, R. Engel, A. Haungs, M. Roth, Czech. J. Phys.
56 (2006) A211.
\bibitem{AugerXmaxMuon1}  A. Aab et al., (Pierre Auger Coll.)
Phys. Rev. D 90 (2014) 012012, erratum: Phys. Rev. D 92 (2015) 019903(E).
\bibitem{AugerMuonExcess2} A. Aab et al., (Pierre Auger Coll.)
Phys. Rev. D 91 (2015) 032003, erratum: Phys. Rev. D 91 (2015) 059901.
\bibitem{qgs} S. Ostapchenko, Nucl. Phys. B (Proc. Suppl.) 151 (2006) 143;
S.~Ostapchenko, Phys. Rev. D 74 (2006) 014026.
\bibitem{sibyll} E.J. Ahn et al., Phys. Rev D 80 (2009) 094003.
\bibitem{Pierog2013b} T. Pierog et al., Phys. Rev. C 92, (2015) 034906.
\bibitem{qgs04} S. Ostapchenko, Phys. Rev. D 83 (2011) 014018.
\bibitem{CIC} J. Hersil et al., Phys. Rev. Lett. 6 (1961) 22; D. M. Edge et al., J.
Phys. A 6 (1973) 1612.
\bibitem{KG} W.D. Apel et al., KASCADE-Grande Collaboration, NIM A 620 (2010) 202.
\bibitem{LDFNKG} K. Kamata, J. Nishimura, Prog. Theor. Phys. Suppl. 6 (1958) 93;
K. Greisen in: J. G. Wilson (Ed.), Progress in Cosmic Ray Physics, Vol. III, J. G.
Wilson (Ed.), North-Holland Publishing Co., 1956.
\bibitem{LDFLR} A.A. Lagutin and R.I. Raikin, Nucl. Phys. B (Proc. Suppl.) 97 (2001)
274.
\bibitem{CORSIKA} D. Heck et al., Report FZKA 6019, Forschungszentrum Karlsruhe,
Germany (1998).
\bibitem{DanielThesis} Daniel Fuhrmann, \textit{KASCADE-Grande measurements of energy
spectra for elemental groups of cosmic rays}, Ph.D. thesis, University of Wuppertal,
Germany (2012).
\bibitem{fluka} A.~Fass\`o et al., Report CERN-2005-10, INFN/TC-05/11, SLAC-R-773
(2005).
\bibitem{Geant} R. Brun, F. Carminati, GEANT-detector description and simulation
tool, CERN Program Library Long Writeup, 1993.
\bibitem{KGPRD} W.D. Apel et al., KASCADE-Grande Collaboration, Phys. Rev. D 87 (2013)
081101(R).
\bibitem{KGNchNmu} W.D. Apel et al., KASCADE-Grande Collaboration, Astrop. Phys.
36 (2012) 183.
\bibitem{Ostap2013} S. Ostapchenko, EPJ Web of Conferences 52 (2013) 02001.
\bibitem{Pierog2013icrc} T. Pierog and D. Heck, in: Proc. 33rd ICRC, Rio de Janeiro,
Brazil (2013) \#icrc163; T. Pierog, J. Phys.: Conf. Ser. 409 (2013) 012008.
\bibitem{Grieder} P. K. F. Grieder, Extensive air showers: High energy
phenomena and astrophysical aspects, a tutorial, reference manual and
data book, Vol. I, Springer, 2011 edition.
\bibitem{Hoerandel} W.D. Apel et al., KASCADE-Grande Collaboration, Phys. Rev.
D 80 (2009) 022002.
\bibitem{Akeno}  M. Honda et al., Phys. Rev. Lett. 70 (1993) 525.
\bibitem{Sugar} C.J. Bell et al., J. Phys. A: Math., Nucl. Gen. 7 (1974) 990.
\bibitem{Ave01} M. Ave et al.,  Proc. 27th ICRC, Hamburg, Germany (2001) 381.
\bibitem{sibyll2.3} F. Riehn, R. Engel, A. Fedynitch, T. K.Gaisser and T. Stanev,
in: Proceedings of Science, PoS (ICRC2015) 558, Proc. 34rd ICRC, The Hague, Netherlands.
\bibitem{Cazon2} L. Cazon et al.,  Astropart. Phys. 36 (2012) 211.
\bibitem{KGHadron2} W.D. Apel et al., KASCADE-Grande Collaboration, Phys. Rev. D
80 (2012) 022002; J.R. H\"orandel, J. Phys. G: Nucl. Part. Phys. 29 (2003) 2439.
\bibitem{MTD2} W.D. Apel et al., KASCADE-Grande Collaboration, Astrop. Phys. 34
(2011) 476.
\bibitem{MTD4}  P. Luczak et al. KASCADE-Grande Collaboration, in: Proceedings of 
Science, PoS (ICRC2015) 386, Proc. 34rd ICRC, The Hague, Netherlands.
\bibitem{KASCADE} T. Antoni et al., KASCADE Collaboration, NIM A 513 (2003) 490.
\bibitem{MTD2005} J. Zabierowski et al. KASCADE-Grande Collaboration, in: Proc. 
29th ICRC, Vol. 6 (2005) 357, Pune, India.
\bibitem{Drescher03} H.J. Drescher and G.R. Farrar, Astrop. Phys. 19 (2003) 235.
\bibitem{Engel01} R. Engel, Rapporteur talk: Particle and interaction physics,
 in: Proc. 27th ICRC, Hamburg, Germany (2001), p. 181.
\bibitem{Haungs01} A. Haungs et al. (KASCADE Coll.), The primary 
 energy spectrum of cosmic rays obtained by muon density measurements at 
 KASCADE, in: Proc. 27th ICRC, Hamburg, Germany (2001), p. 63.
\bibitem{Yakutsk} A.V. Glushkov, et al., Yakutsk EAS array Coll., JETP Lett. 87 
(2008) 190.
\bibitem{AugerMuonExcess}  A. Yushkov, et al., Pierre Auger Coll., Eur.
Phys. J. Web. Conf. 53 (2013) 07002; L. Nellen et al., Pierre Auger Coll.,
J. Phys.: Conf. Ser. 409 (2013) 012107.
\bibitem{MIA} T. Abu-Zayyad et al., Phys. Rev. Lett. 84 (2000) 4276. 
\bibitem{EAS-MSU} Y.A. Fomin, et al., Astrop. Phys. 92 (2017) 1.
\bibitem{ICETOP} J. G. Gonzalez et al., IceCube Coll., J. Phys.: Conf. Ser. 718 
  (2016) 052017. 
\bibitem{KGHadron} W.D. Apel et al., KASCADE-Grande Collaboration, J. Phys. G: Nucl.
Part. Phys. 36 (2009) 035201.
\bibitem{EGS4} W.R. Nelson, H. Hirayama and D.W.O. Rogers, Report SLAC (1985) 265. 
\bibitem{Totem}  G. Antchev et al., TOTEM Coll., Europhys. Lett. 96 (2011) 21002;
101 (2013) 21002; 101 (2013) 21003; 101 (2013) 21004; Phys. Rev. Lett. 111 
 (2013) 012001.
\bibitem{Atlas}  G. Aad, et al., ATLAS Coll., Nucl. Phys. B 889 (2014) 486.
 M. Aaboud et al., ATLAS Coll., Phys. Lett. B 761 (2016) 158.
\bibitem{LambdaNe} T. Antoni et al., Astrop. Phys. 19 (2003) 703.
\bibitem{PierogICRC15} T. Pierog, et al., in: Proceedings of Science, 
PoS (ICRC2015) 337, Proc. 34rd ICRC, The Hague, Netherlands.
\bibitem{Ostap2016} S. Ostapchenko, Phys. Rev. D 93 (2016) 051501(R).
\bibitem{epos199} T. Pierog et al., Report FZKA 7516, Forschungszentrum Karlsruhe,
Germany (2009) 133.
\bibitem{Allen13} J. Allen et al., EPJ Web of Conferences 53 (2013) 01007.
\bibitem{Ulrich11} R. Ulrich et al., Phys. Rev. D 83 (2011) 054026.
\bibitem{NA61}  A. E. Herv\'e for the NA61/SHINE Collaboration, in: Proceedings
 of Science, PoS (ICRC2015) 330, Proc. 34rd ICRC, The Hague, Netherlands.
\bibitem{Zech} G. Zech, Comparing Statistical Data to Monte Carlo Simulation 
- Parameter Fitting and Unfolding, DESY 95-113 (1995);
G. Bohm, G. Zech, NIMA 691 (2012) 171.
\bibitem{KGPRL} W.D. Apel et al. KASCADE-Grande Collaboration, Phys. Rev. Lett.
107 (2011) 171104.
\bibitem{KGunfold} W.D. Apel et al. KASCADE-Grande Collaboration, Astrop. Phys.
47 (2013) 54.
\bibitem{KGMudensities}V. de Souza
et al., KASCADE-Grande Collaboration, in: Proc. 32nd ICRC, Vol. 1580 1/11 
(2011) 295, Beijing, China.

\end{thebibliography}
\end{document}